\documentclass[11pt]{article}
\usepackage{cite}
\begin{document}
\begin{flushright}
gr-qc/0412114\\
\end{flushright}
\vskip 2cm

\begin{center}
{\Large {\bf Cosmological perturbations of brane-induced gravity and the vDVZ
 discontinuity
 on FLRW space-times}}\\
\vskip 1.5cm
C\'edric Deffayet\footnote{deffayet@iap.fr}\\
{\it IAP/GReCO, 98bis Boulevard Arago, 75014 Paris, France \\
and \\
 F\'ed\'eration de recherche APC, Universit\'e  Paris VII,\\
2 place Jussieu - 75251 Paris Cedex 05, France}
\end{center}
\vskip 1cm
\noindent
\begin{center}
{\bf Abstract}
\end{center}
\vskip .3cm
We investigate the cosmological perturbations of the brane-induced (Dvali-Gabadadze-Porrati) model  which exhibits a van Dam-Veltman-Zakharov (vDVZ) discontinuity when linearized over a Minkowski background. We show that the linear brane scalar cosmological perturbations over an arbitrary Friedmann-Lema\^{\i}tre-Robertson-Walker (FLRW) space-time have a well defined limit when the radius of transition between 4D and 5D gravity is sent to infinity with respect to the background Hubble radius. This radius of transition plays for the 
brane-induced gravity model a r\^ole equivalent to the Compton wavelength of the graviton in a Pauli-Fierz theory, as far as the vDVZ discontinuity is concerned. This well defined limit is shown to obey the linearized 4D Einstein's equations whenever the Hubble factor is non vanishing. This shows the disappearance of the vDVZ discontinuity for general FLRW background, and extends the previously know result for maximally-symmetric space-times of non vanishing curvature. Our reasoning is valid for matter with simple equation of state such as a scalar field, or a perfect fluid with adiabatic perturbations, and involves to distinguish between space-times with a vanishing scalar curvature and space-times with a non vanishing one. We also discuss the validity of the linear perturbation theory, in particular for those FLRW space-times where the Ricci scalar is vanishing only on a set of zero measure. In those cases, we argue that the linear perturbation theory breaks down when the Ricci scalar vanishes (and the radius of transition is sent to infinity), in a way similar to what has been found to occur around sources on a Minkowski background. 

\pagebreak

\newcommand{\beq}{\begin{eqnarray}}
\newcommand{\eeq}{\end{eqnarray}}
\newcommand{\kc}{\kappa_{(5)}}
\newcommand{\kq}{\kappa_{(4)}}
\newcommand{\kcd}{\kappa^2_{(5)}}
\newcommand{\kcq}{\kappa^4_{(5)}}
\newcommand{\kqd}{\kappa^2_{(4)}}
\newcommand{\Lc}{\Lambda_{(5)}}
\newcommand{\Lq}{\Lambda_{(4)}}
\newcommand{\abd}{\dot{a}_{(b)}}
\newcommand{\abp}{a^\prime_{(b)}}
\newcommand{\ab}{a_{(b)}}
\newcommand{\nbd}{\dot{n}_{(b)}}
\newcommand{\nbp}{n^\prime_{(b)}}
\newcommand{\nb}{n_{(b)}}
\newcommand{\bb}{{(b)}}
\newcommand{\MM}{{(M)}}
\newcommand{\EE}{{({\cal E})}}
\newcommand{\hd}{\dot{H}}
\newcommand{\hdd}{\ddot{H}}
\newcommand{\hddd} {H^{(3)}}
\newcommand{\da}{\dot{a}}
\newcommand{\db}{\dot{b}}
\newcommand{\dn}{\dot{n}}
\newcommand{\dda}{\ddot{a}}
\newcommand{\ddb}{\ddot{b}}
\newcommand{\ddn}{\ddot{n}}
\newcommand{\paBDL}{a^{\prime}}
\newcommand{\pb}{b^{\prime}}
\newcommand{\pn}{n^{\prime}}
\newcommand{\ppa}{a^{\prime \prime}}
\newcommand{\ppb}{b^{\prime \prime}}
\newcommand{\ppn}{n^{\prime \prime}}
\newcommand{\fda}{\frac{\da}{a}}
\newcommand{\fdb}{\frac{\db}{b}}
\newcommand{\fdn}{\frac{\dn}{n}}
\newcommand{\fdda}{\frac{\dda}{a}}
\newcommand{\fddb}{\frac{\ddb}{b}}
\newcommand{\fddn}{\frac{\ddn}{n}}
\newcommand{\fpa}{\frac{\paBDL}{a}}
\newcommand{\fpb}{\frac{\pb}{b}}
\newcommand{\fpn}{\frac{\pn}{n}}
\newcommand{\fppa}{\frac{\ppa}{a}}
\newcommand{\fppb}{\frac{\ppb}{b}}
\newcommand{\fppn}{\frac{\ppn}{n}}
\newcommand{\UU}{\Upsilon}
\newcommand{\fU}{f_\UU}
\newcommand{\CC}{{\cal C}}
\newcommand{\Back}{{(B)}}
\newcommand{\Lin}{{(L)}}
\newcommand{\Bak}{{(B)}}

\section{Introduction}
Brane world models offer several interesting new ways to 
modify the gravitational interaction and mimic a 4D gravity theory 
 from an intrinsically higher dimensional one. This can be achieved assuming that the $(4+n)$ dimensional bulk space-time has a special topology,
 with  e.g. compact extra-dimensions \cite{Arkani-Hamed:1998rs},  or geometry, 
being e.g. a patch of $AdS_5$ space-time \cite{Randall:1999vf}, and leads generically to modifications of the gravitational interaction at short distances. Variations of these models have been considered where gravity was modified at large (cosmological) distances, being mediated by a resonance of massive graviton propagating over an {\it infinite volume} bulk \cite{GRS}. While the first models of this sort suffered from ghost-like modes being related to the presence of a negative-tension dynamical brane (see \cite{Pilo:2000et} and e.g. the review \cite{Rubakov:2001kp}), the Dvali-Gabadadze-Porrati model \cite{DGP}, relying on an other construction, leads to similar properties for gravity, without 
suffering from the same drawbacks. This model has been applied to cosmology with the aim of explaining the observed late time acceleration of the universe by a large distance modification of gravity \cite{Deffayet:2001uy,Fifth,Deffayet:2002sp}, with the exciting perspective that this can also be tested by solar system gravity tests \cite{SOLAR,SOLARBIS}.  
Aside its interesting phenomenology, the Dvali-Gabadadze-Porrati (DGP in the following) model is also interesting from a more abstractly motivated view.
It can serve as a consistent framework to  study various long standing problems associated with {\it massive gravity}. Here we are using this model as a tool to investigate the van Dam-Veltman-Zakharov (vDVZ) discontinuity \cite{Veltman} over a cosmological, but not necessarily maximally symmetric, Friedmann-Lema\^{\i}tre-Robertson-Walker (FLRW) background. The essence of the vDVZ  discontinuity is that the quadratic Pauli-Fierz \cite{Pauli} action for a massive graviton 
on a Minkowski background 
gives different physical predictions (like e.g. for  light bending) from linearized general relativity whatever the smallness of the graviton mass. It is known, however, that this is no longer true on a maximally symmetric non flat
 background \cite{Higushi,Porrati:2000cp}, when the graviton Compton wavelength (squared) is much larger than the background cosmological constant. The same is expected to be true for more general space-times with non vanishing curvature, where some curvature scale is supposed to play the role of the cosmological constant (square-rooted). Here we use some results obtained elsewhere \cite{Deffayet:2002fn,DEF} for brane-world cosmological perturbations to investigate this issue in the DGP model, in which the graviton propagator has the same tensorial structure as in Pauli-Fierz theory. 

We show that the linearized DGP theory over an arbitrary FLRW space-time has a well defined limit when the radius of transition between 4D and 5D gravity is sent to infinity with respect to the background Hubble radius. This radius of transition plays for the DGP model a r\^ole equivalent to the Compton wavelength of the graviton in a Pauli-Fierz theory, as far as the 
vDVZ discontinuity is concerned. This well defined limit is shown to obey the linearized 4D Einstein's equations whenever the Hubble factor is non vanishing. This shows the disappearance of the vDVZ discontinuity for general FLRW background, and extends the above mentioned result for maximally-symmetric space-times. Our reasoning is valid for matter with simple equation of state such as a scalar field, or a perfect fluid with adiabatic perturbations, and involves to distinguish between brane space-times with a non vanishing scalar curvature and brane space-times with a vanishing one. In the last case, it is shown that in the limiting theory, perturbations are solely supported by the brane motion in an unperturbed Minkowski bulk.
The question of the presence or absence of the vDVZ discontinuity is thus first addressed here in the linearized theory. We also discuss the validity of the linear perturbation theory, in particular for FLRW space-times where the Ricci scalar is vanishing only on a set of zero measure. In those cases it is argued that the linear perturbation theory breaks down where the Ricci scalar vanishes (and the radius of transition is sent to infinity), in a way similar to what has been found to occur around sources on a Minkowski background.

The paper is organized as follows.  In the remaining of this introduction we remind aspects of the vDVZ discontinuity (subsection~\ref{vDVZRecall}), introduce the DGP model (subsection~\ref{DGPmodel}) as well as the solutions for the cosmological background that will be used later (subsection~\ref{VDVZFRW1p14}). 
In section \ref{4Dpert}, we recall some properties of cosmological perturbations in 4D General Relativity (GR in the following).
We then turn to discuss the cosmological perturbations of the DGP model (section~\ref{DGPpert}). First, we introduce the general formalism that we will later  use (subsection~\ref{MOSTGENPERT}), then we apply it to the study of the vDVZ discontinuity as seen in the Newtonian potential exchanged between non relativistic sources (subsection~\ref{[PCGI25p32]}). Next we turn to the case of time dependent perturbations, first by recalling the differential problem associated with this case (subsection~\ref{GCB}), then, by discussing FLRW backgrounds with non vanishing Ricci scalars (subsection~\ref{vDVZ2}), before turning to
 time-dependent perturbations over a Minkowski space-time (subsection~\ref{vDVZ3}). 
We then address the case of a cosmological background with a vanishing scalar curvature, but a non vanishing Hubble factor (subsection~\ref{vDVZ4}), before discussing the generality of some of our previous arguments (subsection~\ref{vDVZ5}). In the following section (section~\ref{SECQUAT}), we discuss the validity range of the linearized theory, in particular for those FLRW space-times where the Ricci scalar vanishes on a set of zero measure. Last, we summarize and  conclude (section~\ref{SECCINQ}).

\subsection{vDVZ discontinuity and Pauli-Fierz theory} \label{vDVZRecall}
The van Dam-Veltman-Zakharov discontinuity \cite{Veltman} was first discussed in the framework of the Pauli-Fierz theory for a massive spin two field \cite{Pauli}. This theory can be defined by the following action
\beq \label{PF}
S^{(\Lq,m)}\equiv S^{(2)}_{\rm EH} + S^{(m)}_{\rm PF} + S_{\rm MC},
\eeq
where $S^{(2)}_{\rm EH}$ is the Einstein-Hilbert action 
\beq
S_{\rm EH} =- \frac{1}{2  \kqd} \int d^4 x \sqrt{-g} \left(R-2 \Lq\right),
\eeq
truncated to quadratic order into a small metric fluctuation $h_{\mu \nu}$, considered as a dynamical field over some background metric $\bar{g}_{\mu \nu}$ solution to Einstein's equations (with $g_{\mu \nu} = \bar{g}_{\mu \nu} + h_{\mu \nu}$ and $\kqd$ being the inverse squared reduced 4D Planck mass), $S^{(m)}_{\rm PF}$ is the Pauli-Fierz mass term, and $S_{\rm MC}$ is the action defining the coupling between the  graviton and matter.
 The Pauli-Fierz mass term is defined by
\beq \label{SPFM}
S^{(m)}_{\rm PF} \equiv \frac{m^2}{8  \kqd} \int d^4x \sqrt{-\bar{g}}   \;
h_{\mu \nu} h_{\alpha \beta} \left(\bar{g}^{\mu\nu} \bar{g}^{\alpha\beta} - \bar{g}^{\alpha\mu} \bar{g}^{\beta\nu}\right),
\eeq
where $m$ is a mass parameter. On a flat background, this mass term gives a mass to the spin two field $h_{\mu \nu}$ in a ghost-free way. The coupling term is given by
\beq
S_{\rm MC} = \frac{1}{2}\int d^4 x\sqrt{-\bar{g}} \; h_{\mu \nu} T^{\mu \nu},
\eeq
where $T^{\mu \nu}$ is the (conserved) matter energy momentum tensor. Note that by definition the action $S^{(\Lq,0)}$ is just the quadratic action for a massless graviton coupled to matter and deduced from general relativity. 
Wishing to compare the massless ($S^{(\Lq,0)}$) and the massive ($S^{(\Lq,m \neq 0)}$) theory, it is convenient to define 
the amplitude ${\cal A}^{(\Lq,m)}$ due to one particle exchange between two conserved energy momentum tensors $T_{\mu \nu}$ and $S_{\mu \nu}$. This amplitude is given formally by 
\beq \label{AMP}
{\cal A}^{(\Lq,m)} &\equiv& \frac{1}{2}\int d^4 x S^{\mu \nu}(x) h_{\mu \nu}\left(T\right)(x) \\
&=-&\kqd \int d^4 x d^4 x'S^{\mu \nu}(x)\Pi_{\mu \nu \alpha \beta}(x,x')T^{\alpha \beta}(x'),
\eeq
where $h_{\mu \nu}\left(T\right)$ is the graviton field generated by $T_{\mu \nu}$, and $\Pi_{\mu \nu \alpha \beta}$ is the graviton propagator.

Let us first consider the case where the background metric 
$\bar{g}_{\mu \nu}$ is the Minkowski metric $\eta_{\mu \nu}$ and the cosmological constant $\Lq$ vanishes. The dramatic observation made in refs. \cite{Veltman} is that, in this case,  the two theories $S^{(0,0)}$ and $S^{(0,m \neq 0)}$ give different physics whatever the smallness of the mass parameter $m$.
Indeed, in momentum space, one has  ($p^\mu$ being the 4-momentum)
\beq
p^2 \;  \Pi_{\mu \nu \alpha \beta}^{(m=0)} &=& \frac{1}{2} \eta_{\mu \alpha} \eta_{\nu \beta} + \frac{1}{2} \eta_{\mu \beta} \eta_{\nu \alpha} - \frac{1}{2} \eta_{\mu \nu}\eta_{\alpha \beta}+ \cdots \label{propzero}\\
(p^2 + m^2)\;  \Pi_{\mu \nu \alpha \beta}^{(m\neq 0)} &=& \frac{1}{2} \eta_{\mu \alpha} \eta_{\nu \beta} + \frac{1}{2} \eta_{\mu \beta} \eta_{\nu \alpha} - \frac{1}{3} \eta_{\mu \nu}\eta_{\alpha \beta}+ \cdots \label{propmass}
\eeq
where momentum-dependent (and gauge-dependent in the massless case) terms have been omitted in the right hand side. The difference in the coefficient of the third  term in the right hand side of the above expressions is responsible for the discontinuity. Note indeed that in the massive case, this coefficient is mass independent. Considering e.g. non relativistic sources separated by a distances sufficiently small with respect to the graviton Compton wavelength, the amplitude due to the exchange of a massive graviton is given approximately by taking the $m \rightarrow 0$ limit in ${\cal A}^{(0,m\neq 0)}$, that we note  ${\cal A}^{(0,m \rightarrow 0)}$. One finds
\beq \label{AMPLIT}
{\cal A}^{(0,m\rightarrow 0)} = \frac{4}{3} {\cal A}^{(0,m=0)},
\eeq
so that the massive amplitude stays different from the massless one, whatever the smallness of the graviton mass. This translates into a similar discrepancy in the non relativistic potential between the two same sources, the potential of the massive theory being larger by a factor 4/3. This extra attraction can be attributed to the exchange of the helicity zero polarization of the massive graviton (which has 3 more polarizations than the massless one).
This difference can be nullified by redefining the Newton constant of the massive theory with respect to the massless one,
assuming, e.g., that one measures the Newton constant by some Cavendish experiment (indeed, if one does not do such a rescaling, the Newton constant of the massive theory would be given by  $4/3 \times \kqd/8\pi = \kqd/6\pi$). However, with such a rescaling, the discontinuity will then reappear in other observables, like the light bending. The latter will then be 25\% smaller in the massive case than in the massless one, which is much too large to be compatible with current measurements of the light bending by the sun \cite{Veltman}.

The situation is however very different when the background space-time is non flat. Indeed taking $\bar{g}$ to parametrize a de Sitter or anti de Sitter space-time with a cosmological constant $\Lq$, one finds that the previously defined amplitude ${\cal A}$ takes the form \cite{Porrati:2000cp}
\beq
{\cal A}^{(\Lambda_{(4)},m)} &=& - \kqd \int S^{\mu \nu} \left(\Delta_L^{(2)} + m^2 - 2\Lq\right)^{-1} T_{\mu \nu} \nonumber \\
&& + \kqd \int S(-\nabla^2+m^2 -2 \Lq)^{-1} \left(\frac{1}{2} + \frac{m^2}{2(2\Lq-3m^2)}\right) T, \label{INTE}
\eeq
where $\Delta_L^{(2)}$ is the Lichnerowicz operator acting on spin-2 symmetric tensors \cite{LICHNE}. Setting $\Lq$ to zero into this expression, one recovers the previous results for a Minkowski background. However, in contrast to the previous case, one has 
\beq
{\cal A}^{(\Lq \neq 0, m \rightarrow 0)} = {\cal A}^{(\Lq \neq 0, m = 0)}.
\eeq
So that the discontinuity disappears on a maximally symmetric background with a non vanishing cosmological constant \cite{Higushi,Kogan1,Porrati:2000cp}\footnote{However the theory is non unitary for $\Lq >0$ and $m^2 < 2  \Lq/3$ \cite{Higushi}.}. This shows that the limits  $m\rightarrow 0$
 and $\Lq \rightarrow 0$ do not commute; however the amplitude ${\cal A}^{(\Lq,m)}$ goes smoothly toward ${\cal A}^{(\Lq,0)}$ when one lets $m^2 \Lq^{-1}$ go to zero.

As was alluded to here-above, the presence of the discontinuity can be attributed to the exchange of the scalar polarization (helicity 0) of the massive graviton. The latter is coupled to the trace of the energy momentum tensor of a conserved source. When the background is Minkowski, this coupling is mass independent and remains in the limit where the graviton mass is sent to zero. However, on a maximally symmetric space-time with non vanishing curvature, the helicity zero coupling becomes proportional to the graviton mass and disappears in the zero mass limit, allowing to recover the massless result. A similar disappearance can be expected to happen in more general cases where the background space-time has non vanishing curvature \cite{Deffayet:2002uk,Arkani-Hamed:2002sp}. This is in particular supported by the study of Schwarzschild-like solutions where it has been argued some time ago by A.Vainshtein \cite{Arkady} that the vDVZ discontinuity could disappear non perturbatively, i.e. in the full exact solution of the theory. Indeed, as noted by Vainshtein, the Schwarzschild radius enters in combination with the graviton mass at the second non trivial order in the perturbation theory to open the possibility for a non perturbative recovery of the short distance behavior the standard Schwarzschild solution. In the case of a maximally symmetric background with non vanishing curvature, however, there is already a length scale (besides the graviton Compton length) at the first order of perturbation theory, the radius of curvature of the background space-time, and, as we saw above, this combines with the graviton mass to lead to a disappearance of the vDVZ discontinuity already at the linear level. 

One thus would wish to study the issue of the disappearance of the vDVZ discontinuity for general curved backgrounds. This is however  plagued by possible inconsistencies. Indeed, the background metric $\bar{g}_{\mu \nu}$, discussed so far, was introduced as a completely extraneous field with respect to the ``graviton'' $h_{\mu \nu}$, and one would need to have a fully non linear theory of gravity which linearized action would be given by (\ref{PF}). However there is no known non pathological such a theory. E.g. it has been shown \cite{BD} that starting from the full Einstein-Hilbert action, and going beyond the quadratic level in the expansion for the kinetic term, causes a ghost-like sixth degrees of freedom to propagate.
Other non linear completion using two metrics (The so called {\it strong 
gravity} or {\it bi-gravity} theories \cite{Isham:gm,Damour:2002ws}) have also been shown to suffer from the same drawbacks \cite{Damour:2002ws}. One could also think to start from a higher dimensional theory and truncate it in some clever ways to retain a finite number of massive graviton Kaluza-Klein states, such truncations are however also pathological \cite{Duff:ea}\footnote{Note also that the non perturbative disappearance \`a la Vainshtein of the vDVZ discontinuity has also been analyzed in some of the above mentioned non linear completion of action (\ref{PF}) with a negative result \cite{Porrati:2002cp,Damour:2002gp}}.

Here we would like to concentrate on a non linear theory which exhibit a vDVZ discontinuity  on flat backgrounds.  This is the brane world model of Dvali-Gabadadze-Porrati (DGP) \cite{DGP} that we will introduce in more details in the next subsection. So far, it has not been found to suffer from the same problems as the other non linear completions we mentioned, and in addition, it has known exact cosmological solutions \cite{Deffayet:2001uy}. The aim of this paper is
 to use those solutions as a consistent background to study the issue of the disappearance of the vDVZ discontinuity over cosmological space-times which are not maximally symmetric.

\subsection{DGP model} \label{DGPmodel}
The DGP model \cite{DGP} we are considering is a 5D brane-world model 
with bulk gravitational action 
\beq \label{S5}
S_{(5)}=-\frac{1}{2 \kcd} \int d^5X \sqrt{g_{(5)}} R_{(5)},
\eeq
where   $R_{(5)}$ is the 5D Ricci scalar, 
and $\kcd$ is the inverse  third power of the reduced 5D Planck mass. To account for the brane, one adds to this action  a term taking care of brane-localized fields given by 
\beq \label{S4}
S_{(4)}=\int d^4 x \sqrt{g_{(4)}} {\cal L},
\eeq
where ${\cal L}$ is a Lagrangian density given by 
\beq \nonumber
{\cal L} = {\cal L}_{(M)} - \frac{1}{2 \kqd} R^{(4)}.
\eeq
In the above expression, ${\cal L}_{(M)}$ is a Lagrangian for brane localized matter and $R^{(4)}$ is the Ricci scalar of the induced metric $g^{(4)}_{\mu \nu}$ on the brane defined by 
\beq
\label{induite}
g^{(4)}_{\mu \nu} = \partial_\mu X^A \partial_\nu X^B g^{(5)}_{AB}, 
\eeq $X^A(x^\mu)$ are defining the brane position, 
where $X^A$ are bulk coordinates, and 
$x^\mu$  coordinates along the brane world-volume\footnote{Throughout this article, we will adopt the following  convention for indices:
upper case Latin letters $A,B,...$ will denote 5D indices: $0,1,2,3,5$; 
Greek letters  $\mu,\nu,...$ 
 will denote 4D indices parallel to the brane: $0,1,2,3$;
lower case Latin letters from the middle of the
alphabet: $i,j,...,$ will denote space-like 3D indices parallel to the brane: $1,2,3$.}. We also implicitly include in the action a suitable Gibbons-Hawking term \cite{Gibbons:1976ue} for the brane. It is always possible to choose a, so called Gaussian Normal (referred to as GN in the rest of this work), coordinate system where the line element 
can be put in the form 
\beq \label{GN}
ds^2 = dy^2 + g^{(4)}_{\mu \nu} dx^\mu dx^\nu, 
\eeq
and the brane is the hyper-surface defined by $y=0$.
In this coordinate system the gravitational  equations of motion derived from the sum of actions 
(\ref{S5}) and (\ref{S4}) are simply given by 
\beq \label{einstein}
G^{(5)}_{AB} = \kcd \delta (y) \left(T_{\mu \nu}^\MM - \frac{1}{\kqd}G_{\mu \nu}^{(4)} \right) \delta_A^\mu \delta_B^\nu, 
\eeq
where $G^{(5)}_{AB}$ is the 5D Einstein tensor, 
$G^{(4)}_{\mu \nu}$ is the 4D Einstein tensor built from the induced metric (\ref{induite}), and $T_{\mu \nu}^\MM$ is the energy-momentum tensor of the brane localized matter fields entering into ${\cal L}_\MM$.  

Before discussing the DGP model defined by the equations of motion (\ref{einstein}), or the sum of actions (\ref{S5}) and (\ref{S4}), let us first turn to a scalar toy model, proposed in the original reference \cite{DGP}, which captures some key features of the full DGP model. It can be defined by the action for the suitably normalized scalar field $\phi$,
\beq
S_{\phi}=  \int d^4x dy \left\{ \frac{1}{\kcd}\partial_A \phi \partial^A \phi + \frac{1}{2}\delta(y) J_{(4)} \phi + 
\frac{1}{\kqd}  \delta(y) \partial_\mu \phi \partial^\mu \phi \right\}. 
\eeq 
This model is simply one for a scalar field in a 5D flat bulk with a brane localized (in $y=0$) kinetic term added for it, and a brane localized source term $J_{(4)}$. The equations of motion derived from this action read 
\beq \label{GREENSCA}
\left(\frac{1}{\kcd} \partial^A\partial_A + \frac{1}{\kqd} \delta(y) \partial_\mu \partial^\mu\right)\phi = \frac{1}{2}\delta(y) J_{(4)}.
\eeq
In this model, the interaction potential between two static sources, separated by a distance $r$, interpolates between a 4D $1/r$ behavior at small distances and a 5D $1/r^2$ behavior at large distances, as shown in reference \cite{DGP}. The crossover distance $r_c$ between the two regimes being is given by 
\beq
r_c = \frac{\kcd}{2\kqd}.
\eeq

The same behavior was found in ref. \cite{DGP} to hold for the gravitational potential between two sources put on the brane in the theory defined by the sum of actions (\ref{S5}) and (\ref{S4}), whenever the background bulk and brane space-times were both taken to be Minkowski space-times. However it was also found there that, from a 4D point of view, gravity is mediated by a continuum of massive Kaluza-Klein modes, with no normalizable zero mode entering into the spectrum. This being a consequence of the bulk being flat and infinite. As a consequence the tensorial structure of the graviton propagator was shown to be the one of a massive graviton and the model exhibits a vDVZ discontinuity. E.g. the  Fourier transform of the amplitude, defined as in (\ref{AMP}),  between two conserved sources $T_{\mu \nu}$ and $S_{\mu \nu}$ on the brane (which position is defined by $y=0$), takes the form \cite{DGP}
\beq \label{AMPLITDGP}
\check{h}_{\mu \nu} (p,y=0) \check{S}^{\mu \nu}(p)  \propto \frac{\check{T}^{\mu \nu}\check{S}_{\mu \nu} - \frac{1}{3}
\check{S}^\mu_\mu\check{T}^\nu_\nu}{p^2 + r_c^{-1} p}, 
\eeq  
where inverse hats denotes Fourier-transformed quantities (and one has gone to the Euclidean space, so that $p$ is simply the square root of the four momentum square). One notes above the factor $1/3$ on the numerator similar to the one seen in the propagator (\ref{propmass}), while the mass of the graviton appearing in 
(\ref{propmass}) is here replaced by the ``momentum dependent mass'' $ r_c^{-1} p$, so that $r_c^{-1}$ plays in the DGP model a r\^ole similar to the mass scale $m$ of the Pauli-Fierz action \footnote{Indeed one can view (\ref{S5}) as a corrective term  in $r_c^{-1}$ to the 4D action (\ref{S4}), in the same way the Pauli-Fierz mass term (\ref{SPFM}) can be considered as a perturbation in $m^2$ of the massless action $S_{EH}^{(2)}$. The vDVZ discontinuity is then stating that this perturbative expansion does not translates in a similar expansion in observables of interest.}. As discussed in the previous subsection, the vDVZ discontinuity would seem to rule out phenomenological applications of DGP model, if one trusts the linear approximation to compute observables such as the light bending from the sun. However, based on the fact that the exact cosmological solutions (found in \cite{Deffayet:2001uy} and recalled in the next subsection) do not exhibit any sign of the vDVZ discontinuity, it has been suggested that the vDVZ discontinuity of the DGP model was indeed an artifact of the perturbation theory \cite{Deffayet:2002uk} following the similar suggestion made by A.~Vainshtein in the framework of Pauli-Fierz theory \cite{Arkady}, and recalled in the previous subsection. This has been studied in different situations by  expansions going beyond the linear order in the brane bending \cite{Lue:2001gc,Gruzinov:2001hp,Porrati:2002cp,Tanaka:2003zb,Gabadadze:2004iy}. Those analysis, mostly concentrated on the case of 
spherically symmetric solutions on the brane, confirmed so far the original suggestion made in ref. \cite{Deffayet:2002uk}:  for a spherically symmetric metric on the brane, the solution given by the linearized theory breaks down below a non perturbative distance, given by \footnote{Note however that all the spherically symmetric solutions studied so far are approximate solutions.
This has several aspects. First, the recent work \cite{Gabadadze:2004iy} suggests that the linearized theory is never a good approximation of the exact non perturbative solution of the brane (even if this work also finds the appearance of the scale (\ref{strongcoupl}) as well as a non perturbative recovery of the massless tensorial structure below that scale). Second,
 the fact that the exact spherically symmetric solution is not known still allows a loophole to all analysis done so far. One could indeed argue that pathologies are arising in the exact solution, similar to those found in ref \cite{Damour:2002gp} in the context of bi-gravity. However the situation of the DGP model with respect to this issue is as good (or as bad) as in any other brane world models (like the Randall Sundrum model \cite{Randall:1999vf}) where the exact (stable) Schwarzschild-type  solution on the brane, corresponding to the perturbative analysis, is not known.} 
\beq \label{strongcoupl}
r_v = \left(r_c^2 r_S\right)^{1/3}. 
\eeq
This scale can be coined from the equivalent one found by A.~Vainshtein to appear for a Pauli-Fierz model \cite{Arkady}. It depends on a curvature scale of the spherically symmetric metric, namely the 4D Schwarzschild radius $r_S$. For distances much lower than $r_v$ it has been found that the spherically symmetric solution on the brane  is close to a usual 4D Schwarzschild metric (with 4D, massless, ``tensorial structure'') so that there is no more discontinuity. In particular, one sees that $r_v$ diverges as $r_c$ goes to infinity (this is similar to sending $m$ to zero in Pauli Fierz action), the 4D parameters ($\kqd$, and the mass of the source) being fixed.
This disappearance of the vDVZ discontinuity  can be attributed to the fact that one mode entering into the gravitational exchange, and related to the brane bending (or the brane extrinsic curvature), has a cubic interaction which becomes important at distances lower than $r_v$, as discussed in detail in \cite{Deffayet:2002uk}. 
This ``strong coupling'' is a blessing from the classical point of view, since it enables to recover the right tensor structure. A debate is  currently going on on the quantum consequences of this strong coupling \cite{DEBATE,Nicolis:2004qq}. This will not matter for us, as far as the conclusions of this paper are concerned, since our discussion is here purely classical. Moreover we will mostly be concerned here with the linear perturbation theory. And, following the discussion of the previous subsection as well as what is suggested by the study of spherically symmetric solutions, ask whether already at the linear level, the vDVZ discontinuity disappears in the DGP model on cosmological background, when we let $r_c$ go to infinity and keep the parameters of the $4D$ part of the action, (\ref{S4}), fixed. 
We will come back on this ``strong coupling'' issue in section \ref{VDVZFRWPAR28}.

We thus need to look at the DGP   equations of motion (\ref{einstein}) linearized over a cosmological background. That is to say we need to look at the so-called {\it cosmological perturbations} of the DGP model. We will only consider the case of scalar perturbations (as seen from an observer on the brane) since it is for those perturbations that the discontinuity is expected to show up, being related to the helicity-zero mode of the graviton.  In the next subsection, we discuss some properties of the cosmological background we are interested in.   

\subsection{Cosmological background for DGP model}
It is convenient to discuss the cosmological background in a so-called Gaussian Normal coordinate system with respect to the brane, where the background metric takes the form 
\beq \label{backmet}
ds^2_{(5)} = -n^2(t,y) dt^2 + a^2(t,y) \delta_{ij} dx^i dx^j + dy^2, 
\eeq
where the 3D metric $\delta_{ij}$ is a flat Euclidean metric (we will only consider in this paper the case of a spatially flat universe), and the brane sits at $y=0$. One can further choose a time parametrization such that the function $n$ is set to one on the brane (that we note $n_{(b)} =1$, where the index $(b)$ means here, and in the following, that the corresponding quantity is taken on the brane). In this case  the induced metric (\ref{induite}) is simply given by 
\beq \label{FLRW}
ds^2_{(4)} = -dt^2 + a^2_{(b)} \delta_{ij} dx^i dx^j,
\eeq
(where $a_{(b)} \equiv   a(t,y=0)$) and 
is of Friedmann-Lema\^{\i}tre-Robertson-Walker (FLRW) form. Considering comoving observers to sit at fixed comoving coordinates $x^i$ on the brane, 
$t$ is then simply the cosmological time on the brane. 
Accordingly with the symmetries of (\ref{backmet}) the 
brane localized matter energy momentum tensor is taken of the form 
\beq \nonumber
S^{\mu}_{\nu} = \delta(y) \mbox{diag} \left(-\rho_\MM,P_\MM,P_\MM,P_\MM \right).
\eeq
\label{VDVZFRW1p14}
Solving the bulk Einstein's equations one gets then the expressions for $n$ and $a$ \cite{BDL,Deffayet:2001uy} in the case of an expanding brane (i.e. when $\dot{a}_\bb$ is strictly positive): 
\beq \label{a}
a(t,y) &=& a_\bb \left\{1- \eta |y| H\right\}, \\
n(t,y) &=& 1 -\eta |y| \frac{\dot{H} + H^2}{H} \label{n}, 
\eeq
 where $H\equiv \dot{a}_\bb/a_{\bb}$ is the brane Hubble parameter, and $\eta$ is the sign\footnote{$\eta$, which takes the values $\pm 1$, is equal to $-\epsilon$ of reference  \cite{Deffayet:2001uy}} of the brane effective energy density $\rho \equiv \rho_{(M)} - 3 H^2/\kqd$. Note that the metric (\ref{backmet}) with $a$ and $n$ given by the above expressions, and $y$ of a definite sign (i.e. considering only the $y > 0$ or $y < 0$ part of the space-time), is simply a reparametrization of 5D Minkowski space-time, with coordinate singularities where $a$ or $n$ are vanishing \cite{Deruelle:2000ge,Deffayet:2001uy}. The time evolution of $a_\bb$ is then obtained by solving the Friedmann equations of the DGP model, given by
\beq \label{fried1}
\dot{\rho}_\MM &=& - 3 H (P_\MM + \rho_\MM),  \\
1 &=& - \frac{\Upsilon}{2} + \sqrt{\frac{\kqd \rho_\MM}{3 H^2} + \frac{\Upsilon ^2}{4}}, \label{fried2}
\eeq
 where we have defined the parameter $\Upsilon$ by 
\beq \label{Upsi}
\Upsilon = \frac{\eta}{H r_c}. 
\eeq
The early time behavior of the cosmology defined by equations (\ref{fried1}) and (\ref{fried2}) is obtained by considering the limit $\Upsilon \rightarrow 0$. In this limit (which means that the Hubble radius $H^{-1}$ is much smaller than the crossover radius $r_c$), equation   (\ref{fried2}) is identical to the usual first Friedmann equation and one recovers standard cosmology (with no sign of the vDVZ discontinuity, as recalled above). At late time, however, one has deviations from standard cosmology,  
 with the exciting possibility provided by the $\eta = -1$ branch to yield an accelerated expansion with vanishing cosmological constant \cite{Deffayet:2001uy,Fifth,Deffayet:2002sp}. Note that, for a fixed $\eta$ and a specified matter equation of state (or for given matter equations of motion), the Cauchy problem associated to the system (\ref{fried1}) and (\ref{fried2}) is well posed and the specification of the initial values of $a_\bb$ and $\rho_\MM$ lead to a unique solution for the scale factor on the brane. In particular, as will turn out to matter for the following discussion, one can choose a matter equation of state (albeit quite unconventional), and initial conditions, such that the brane background Ricci scalar $R^\Bak_\bb$, given by $R^\Bak_\bb \equiv 12 H^2 + 6\dot{H}$, vanishes at all times. 

\section{Scalar cosmological perturbations in 4D general relativity}
Before turning to discuss the cosmological perturbations of the DGP model, let us here first recall some features of standard (4D) cosmological perturbations 
(see e.g. \cite{Mukhanov:1992me} for reviews). We will do so in a way suited for comparison with the brane-world cosmological perturbations that we will discuss thereafter. We consider a 
 given cosmological FLRW background space-time, solution of the Friedmann equations derived from 4D GR. 
\label{4Dpert}
We will work in the longitudinal gauge, where the (4D) linearized line element reads
\beq \label{4Dlinepert}
ds^2 = -(1+2 \Phi) dt^2 + a^2(t)(1-2 \Psi) \delta_{ij} dx^i dx^j, 
\eeq
where $a(t)$ is the background scale factor, $\Phi$ and $\Psi$ are the two left over scalar perturbations of the metric. For simplicity, we only consider a spatially flat cosmological background (so that the here-above $\delta_{ij}$ denotes a 3D flat Euclidean metric).   
The linearized Einstein equations 
\beq \label{4Dein} 
\kqd \delta T^\MM_{\mu \nu} =\delta G^{(4)}_{\mu \nu}, 
\eeq
 then lead to the four scalar equations
\beq
\label{EQM6_4D}
  \kqd \delta \rho_\MM &=&  -6 \frac{\dot{a}}{a} \left( \dot{\Psi} + \frac{\dot{a}}{a} \Phi \right) + \frac{2}{a^2} \Delta \Psi,\\
  \kqd \delta q_\MM &=&  -2 \dot{\Psi} - 2 \frac{\dot{a}}{a} \Phi  \label{EQM4_4D},\\
  \kqd \delta \pi_\MM &=& \frac{\Psi - \Phi}{a^2}  \label{EQM1_4D}, \\ 
 \kqd \delta P_\MM &=&   
 2\left(2  \frac{\ddot{a}}{a} + \frac{\dot{a}^2}{a^2} \right)\Phi + 2 \frac{\dot{a}}{a} \dot{\Phi} +  
2 \ddot{\Psi} + 6 \frac{\dot{a}}{a} \dot{\Psi} + \frac{2}{3 a^2} \Delta \left( \Phi - \Psi\right),
  \label{EQM8_4D}
 \eeq
where $\Delta$ is the Laplacian with respect to the comoving coordinates $x^i$, 
the scalar components $ \delta \rho_\MM, \delta  q_\MM, \delta \pi_\MM, \delta P_\MM$ of the matter energy momentum tensor $T^\MM_{\mu \nu}$ are defined by
\beq
\delta {T}^0_{0 \MM} &=& -\delta {\rho}_\MM, \label {dTM00}\\
\delta {T}^0_{i \MM} &=& \partial_i \delta {q}_{ \MM}, \label{dTM0i}\\
\delta {T}^i_{j \MM} &=& \delta {P}_\MM \delta^i_j +  \left(\Delta^i_j - 
\frac{1}{3} \delta^i_j \Delta  \right)   \delta {\pi}_{ \MM } \label{dTMij},  
\eeq
and are respectively the scalar energy density, momentum, anisotropic stress and pressure matter perturbations.
In the above expression $ \Delta^i_j$ is defined by 
\beq \nonumber
 \Delta^i_j = \delta^{ik} \partial_k \partial_j, 
\eeq
so that one has $\Delta = \Delta^i_i$.
Equations (\ref{EQM6_4D}-\ref{EQM8_4D})
 are respectively the $tt$, $ti$, traceless part of $ij$, and trace of $ij$ components of the scalar part of Einstein's equations  (\ref{4Dein}).
The system (\ref{EQM6_4D}-\ref{EQM8_4D}) does not close in general. 
To close the system and solve for the evolution of cosmological perturbations once initial conditions are provided, one needs to add more equations, 
namely the matter equation of motion, or equation of state.
A first simplifying hypothesis, that we will make in the following, is to assume a vanishing $\delta \pi_\MM$, 
\beq \label{perfect}
\delta \pi_\MM = 0.
\eeq
This occurs in the case of a perfect fluid, but also when the only matter is a scalar field. 
Then equation (\ref{EQM1_4D}) leads to the equality between the two gravitational potential $\Phi$ and $\Psi$,
\beq \label{psiphi}
\Phi = \Psi.
\eeq
The potential $\Phi$ (or $\Psi$)
 can be determined by adding again an extra equation. E.g. in the case of adiabatic perturbations, the matter perturbations follow the equation of state 
\beq \label{adiab}
\delta P_\MM = c_S^2 \delta \rho_\MM,
\eeq 
where $c_S^2$ is the sound velocity.
Equation (\ref{adiab}) can be used to eliminate $\delta P_\MM$ and $\delta \rho_\MM$ from the linearized Einstein equation to yield the
evolution equation for the gravitational potential $\Phi$
\label{PCGI 24 page 10}
\beq  \label{EVOL}
\ddot{\Phi}  + (4 + 3 c_S^2) \frac{\dot{a}}{a} \dot{\Phi} + \left[ 2 \frac{\ddot{a}}{a} + \frac{\dot{a}^2}{a^2} (1 + 3 c_S^2) \right] \Phi - \frac{c_S^2}{a^2} \Delta \Phi = 0,
\eeq 
and all the other relevant quantities can be deduced from the knowledge of $\Phi$ as a function of time, as can be seen from equations (\ref{EQM6_4D}-\ref{EQM8_4D}). 
A similar reasoning holds for a scalar field which energy-momentum perturbations obeys the equation of state 
\beq \label{STATSCA}
\delta P_\MM= \delta \rho_\MM   - \delta q_\MM \left(2 \frac{\ddot{\phi}^\Back}{\dot{\phi}^\Back}+ 6H\right), 
\eeq
where $\phi^\Back$ is the background value of the scalar field. Equation (\ref{STATSCA}) then yields in a similar way as for adiabatic perturbations, an ordinary differential equation for $\Phi$.
Note further that (\ref{EVOL}) is a second order differential equation so that one needs as initial conditions the values of $\Phi$ and $\dot{\Phi}$. This will further specify the initial values of all the matter perturbations thanks to equations (\ref{EQM6_4D}), (\ref{EQM4_4D}), (\ref{perfect}), (\ref{adiab}), and (\ref{psiphi}) which can then be regarded as (initial) constraints.  
Last, we note that when one considers non relativistic matter sources on a Minkowski background (where $a$ is constant with time and is set to one), one finds from equation (\ref{EQM6_4D}) the usual Poisson equation for the gravitational potential, reading with our normalization
\beq \label{Poisson}
\Delta \Phi = \frac{\kqd}{2} \delta \rho_\MM.
\eeq
 
\section{Scalar cosmological perturbations in the DGP model} \label{DGPpert}
\subsection{General formalism and Mukohyama's master variable} \label{MOSTGENPERT}
The most general scalar perturbations of the line element (\ref{backmet}) read 
\beq \label{linepertzero}
g_{AB} = \left(\begin{array}{ccc}
-n^2(1+2 \bar{A}) & a^2 \partial_i\bar{B} & n \bar{A}_y \\
a^2 \partial_i\bar{B} & a^2\left[(1+2 \bar{\cal R})\delta_{ij}+ 2 \partial^2_{ij}\bar{E}\right]&a^2 \partial_i\bar{B}_{y}\\
n \bar{A}_y & a^2\partial_i\bar{B}_{y} &1+2\bar{A}_{yy}
\end{array}\right),
\eeq
out of which one can define four gauge-invariant quantities. Following \cite{Bridgman:2001mc}, we define them as $\tilde{A}$, $\tilde{A}_y$, $\tilde{A}_{yy}$, and $\tilde{\cal R}$ given by 
\beq
\tilde{A} &=& \bar{A} - \frac{1}{n}\left(\frac{a^2 \bar{\sigma}}{n}\right)^. + \frac{n^\prime}{n} a^2 \bar{\sigma}_y, 
\label{AGIV} \\ \label{AYGIV}
\tilde{A}_y &=& \bar{A}_y + \frac{(a^2 \bar{\sigma}_y)^.}{n}+\frac{(a^2 \bar{\sigma})^\prime}{n} - 2 \frac{n^\prime}{n^2}a^2 \bar{\sigma}, \\ \label{AYYGIV}
\tilde{A}_{yy} &=& \bar{A}_{yy} + (a^2 \bar{\sigma}_y)^\prime, \\ \label{RGIV}
\tilde{\cal R} &=& \bar{\cal R} + a a^\prime \bar{\sigma}_y - \frac{a \dot{a}}{n^2} \bar{\sigma},
\eeq
where  
a dot means a derivative with respect to $t$, a prime means a derivative with respect to $y$, and  $\bar{\sigma}$ and $\bar{\sigma}_y$ are given by 
\beq
\bar{\sigma} &\equiv& -\bar{B} + \dot{\bar{E}}, \nonumber \\
\bar{\sigma}_y & \equiv& -\bar{B}_y +\bar{E}^\prime.  \nonumber
\eeq
In an analogous way, one can define a gauge invariant perturbed brane position 
 $\tilde{\xi}$ as \cite{Bridgman:2001mc} 
\beq \label{GIVPOS}
\tilde{\xi} = \bar{\xi} - a^2 \bar{\sigma}_y,
\eeq
where $\bar{\xi}$ denotes the linear perturbation of the brane position around the background value $y=0$. The gauge invariant variables $\tilde{A}, \tilde{A}_y, \tilde{A}_{yy}, \tilde{R}$, as well as $\tilde{\xi}$,
are invariant under the scalar gauge transformations
\beq  \label{5Dt}
t &\rightarrow& t + \delta t, \\
x^i &\rightarrow& x^i + \delta^{ik} \partial_k \delta x ,\label{5Dx}\\
y &\rightarrow& y + \delta y, \label{5Dy}
\eeq
where $\delta t$, $\delta x$ and $\delta y$ are arbitrary functions of $t$, $x^i$, and $y$. \label{muko}
The next step would be to introduce the decomposition (\ref{linepertzero}) into Einstein's equations. However it will turn out more convenient for us to use rather Mukohyama's Master variable formalism \cite{Mukohyama:2000ui}, that we now recall. This can be done here, since the bulk background cosmological space-time of the DGP model is a slice of 5D Minkowski space-time.

Mukohyama showed indeed that all the linearized scalar  Einstein's equations over a maximally symmetric background bulk can be conveniently
solved introducing a master variable $\Omega$
 which obeys a PDE in the bulk, the master equation.
The latter, when $\Omega$ has a non trivial  dependence  in the comoving coordinates $x^i$, reads in the  GN coordinate system (\ref{backmet}) 
\beq 
\label{FQ1}
\left(\frac{\Omega^\cdot}{na^3}\right)^\cdot - 
\frac{n \Delta \Omega}{a^5} - \left(\frac{n \Omega^\prime
}{a^3} \right)^\prime=0.
\eeq
In the rest of this article, we will implicitly consider all the perturbations as Fourier transformed  
with respect to the $x^i$s, in order to do a mode by mode analysis.
In particular  (\ref{FQ1}) can be rewritten as 
\beq \label{MASTER2D}
{\cal D}_\Delta \Omega =0, 
\eeq
where ${\cal D}_\Delta$ is a second order hyperbolic differential operator acting on $y$ and $t$ dependent functions (in the GN system), and  $\Delta$ 
is understood to be replaced by  $-\vec{k}^2$, where $\vec{k}$ is the comoving momentum. 
 Equation (\ref{FQ1}) is then only valid when $\vec{k}^2$ does not vanish \cite{Mukohyama:2000ui}, which is the only case of interest as far as cosmological perturbations are concerned. The gauge invariant scalar perturbations (\ref{AGIV}-\ref{RGIV})
can be expressed in term of $\Omega$ as follows \cite{Mukohyama:2000ui,Bridgman:2001mc}
\beq \label{EQ1}
\tilde{A} &=& -\frac{1}{6a} \left(2 \Omega^{\prime \prime} + \frac{1}{n^2}
 \Omega^{\cdot \cdot} -\frac{\dot{n}}{n^3} {\Omega}^\cdot - \frac{n^\prime}{n} \Omega^\prime \right),
\\ \label{EQ6}
\tilde{A}_y &=& \frac{1}{an} \left({\Omega}^{\cdot \prime} - \frac{n^\prime}{n} {\Omega}^\cdot \right),
\\ \label{EQ4}
 \tilde{A}_{yy} &=& \frac{1}{6a} \left( \Omega^{\prime \prime} +
\frac{2}{n^2} {\Omega}^{\cdot \cdot}  - 2 \frac{\dot{n}}{n^3} {\Omega}^\cdot - 2 \frac{n^\prime}{n} \Omega^\prime \right),
\\ \label{EQ20}
\tilde{{\cal R}} &=& \frac{1}{6a} \left( \Omega^{\prime \prime} - \frac{1}{n^2} {\Omega}^{\cdot \cdot} + \frac{\dot{n}}{n^3} {\Omega}^\cdot + \frac{n^\prime}{n} \Omega^\prime \right).
\eeq

Let us now turn to the brane.
One can choose a gauge where the brane sits in $y=0$ and the perturbed induced metric on the brane has the longitudinal form (\ref{4Dlinepert}) (with the brane scale factor $a_\bb$ replacing $a(t)$). The brane localized matter perturbations are decomposed as in (\ref{dTM00}-\ref{dTMij}), where $T_{\mu\nu}^{(M)}$ is now the energy momentum tensor of the brane localized matter fields.  
We will assume in the rest of this work that the matter anisotropic stress perturbation $\delta \pi_{(M)}$ vanishes. 
The Israel's junction conditions \cite{Israel} together with the master equation (\ref{FQ1}) then enables one to express all the scalar degrees of freedom as seen by an observer on the brane, that is to say $\delta \rho_{(M)}$, $\delta q_{(M)}$, $\delta P_{(M)}$, $\Phi$  and $\Psi$, as 
functions of the master variable $\Omega$ and its derivatives, as follows \cite{Deffayet:2002fn}  
\beq
\Phi &=& \frac{1}{6 a_\bb} \left\{ \frac{\Delta \Omega}{a^2}  \CC^{\Phi}_{\Delta (0,0)}
+ H {\Omega}^\cdot  \CC^{\Phi}_{(1,0)} +{\Omega}^{\cdot \cdot}   \CC^{\Phi}_{(2,0)} + 
 H \Omega^\prime \CC^{\Phi}_{(0,1)} \right\}_\bb  \label{PHI1},\\
\Psi &=& \frac{1}{6 a_\bb} \left\{ \frac{\Delta \Omega}{a^2}  \CC^{\Psi}_{\Delta (0,0)}
+ H {\Omega}^\cdot  \CC^{\Psi}_{(1,0)} +  {\Omega}^{\cdot \cdot}  \CC^{\Psi}_{(2,0)}+ 
H \Omega^\prime  \CC^{\Psi}_{(0,1)}  \right\}_\bb \label{PSI1},\\
\kqd \delta \rho_\MM& =& \frac{H^2}{2 a_\bb} \left\{
 \frac{\Delta \Omega}{a^2}  \CC^\rho_{\Delta (0,0)} +  \frac{\Delta^2 \Omega}{H^2 a^4}  \CC^\rho_{\Delta^2 (0,0)}
+  H {\Omega}^\cdot \CC^\rho_{(1,0)} +   {\Omega}^{\cdot \cdot}  \CC^\rho_{(2,0)} 
+  \frac{\Delta {\Omega}^{\cdot \cdot}}{H^2 a^2}  \CC^\rho_{\Delta (2,0)}  \nonumber \right.\\
&& \left.+ \frac{\Omega^{\cdot \cdot \cdot}}{H} \CC^\rho_ {(3,0)} + 
H \Omega^\prime \CC^\rho_{(0,1)} +  \frac{\Delta \Omega^\prime}{H a^2}  \CC^\rho_{\Delta (0,1)} +  \Omega^{\cdot \prime}   \CC^\rho_ {(1,1)}
\right\}_\bb \label{DR1},\\
\kqd \delta q_\MM& =& \frac{H}{6 a_\bb} \left\{
 \frac{\Delta \Omega}{a^2}  \CC^q_{\Delta (0,0)} +
 H {\Omega}^{\cdot}  \CC^{q}_{(1,0)} + \frac{\Delta {\Omega}^\cdot}{H a^2} \CC^{q}_{\Delta (1,0)}
+ {\Omega}^{\cdot \cdot}   \CC^{q}_{(2,0)} +  \frac{\Omega^{\cdot \cdot \cdot}}{H} \CC^q_ {(3,0)} \nonumber \right.\\
&&  \left.
+  H \Omega^\prime \CC^q_{(0,1)} +  \Omega^{\cdot \prime}   \CC^q_ {(1,1)} \right\}_\bb, \label{DQ1}\\
\kqd \delta P_\MM& =& \frac{H^2}{6 a_\bb} \left\{
 \frac{\Delta \Omega}{a^2}  \CC^P_{\Delta (0,0)} + H {\Omega}^\cdot  \CC^{P}_{(1,0)} + \frac{\Delta {\Omega}^\cdot}{H a^2} \CC^{P}_{\Delta (1,0)}
+  {\Omega}^{\cdot \cdot}  \CC^P_{(2,0)}
+  \frac{\Delta \ddot{\Omega}}{H^2 a^2}  \CC^P_{\Delta (2,0)}  \nonumber \right.\\
&&  \left. +  \frac{\Omega^{\cdot \cdot \cdot}}{H} \CC^P_ {(3,0)}  + \frac{\Omega^{\cdot \cdot \cdot \cdot}}{H^2} \CC^P_ {(4,0)}
+  H \Omega^\prime \CC^P_{(0,1)} +  \Omega^{\cdot \prime}   \CC^P_ {(1,1)} +  \frac{\Omega^{\cdot \cdot \prime}}{H}   \CC^P_ {(2,1)}
\right\}_\bb, \label{DP1}
\eeq
where the coefficients $\CC$ are given in the appendix  \ref{expressC}, and our normalization is chosen so that they are dimensionless\footnote{Note that the coefficients $\CC$ are given here in full generality for what concerns the sign of $\eta$; while  they are only given for the case $\eta = +1$ in reference \cite{Deffayet:2002fn}.}.
We note here incidentally that despite the assumed vanishing of the matter anisotropic stress, equations (\ref{PHI1}) and (\ref{PSI1}) show that $\Phi$ and $\Psi$ are not equal in general.  In fact, in the gauge considered here, the difference between $\Phi$ and $\Psi$ is related to the gauge invariant brane position by\footnote{This can be easily seen from the formulae given in ref. \cite{Deffayet:2002fn}.}  
\beq \label{POSPHIPSI}
\tilde{\xi} = r_c \left(\Phi-\Psi\right).
\eeq

Let us first apply this formalism to the case of perturbations over a Minkowski brane, and show how the vDVZ discontinuity does appear there in the Newtonian potential between non relativistic sources. 

\subsection{vDVZ discontinuity on Minkowski background, as seen in the Newtonian potential} 
\label{[PCGI25p32]}
Setting $H$ to zero, $a_\bb$ to one,  in equations (\ref{PHI1}-\ref{DP1}), and keeping $r_c$ finite, one gets the following expressions \label{vDVZ1}
for the gravitational potentials, the matter density, momentum, and pressure perturbations over a Minkowski brane as function of the master variable
\label{[PCGI26p36]}
\beq
\Phi &=& \frac{1}{3} \Delta \Omega_\bb  - \frac{1}{2} \Omega^{\cdot \cdot}_\bb  \label{HOPhiDGP},\\
\Psi &=& \frac{1}{6}  \Delta \Omega_\bb \label{HOPsiDGP}, \\
\kqd \delta \rho_\MM &=& \frac{1}{2} \left\{\Delta^2 \Omega  - \Delta \Omega^{\cdot \cdot} + \frac{1}{r_c} \Delta \Omega^\prime  \right\}_\bb,
\label{HOdrDGP}\\
\kqd \delta q_\MM &=& -\frac{1}{2} \left\{\Delta\Omega^{\cdot}  -  \Omega^{\cdot \cdot \cdot} + \frac{1}{r_c} \Omega^{\cdot \prime}  \right\}_\bb,
\label{HOdqDGP}\\
\kqd \delta P_\MM &=& \frac{1}{2} \left\{\Delta \Omega^{\cdot \cdot}  -  \Omega^{\cdot \cdot \cdot \cdot} + \frac{1}{r_c} \Omega^{\cdot \cdot \prime}  \right\}_\bb. \label{HOdPDGP}
\eeq
\label{[PCGI26p29]}
As we noted above, $\Phi$ is not in general equal to $\Psi$,
 this remaining true 
 even if one sends $r_c$ to infinity
 ({\it after} sending H to zero) as will be seen more explicitly below. It is the trace of \label{[VDVZPAR12]}
 the presence of the  vDVZ
 discontinuity  in the DGP model (and can be translated into the presence of a 
non vanishing anisotropic stress for the Weyl's fluid, which 
remains in the limit where $r_c$ goes to infinity on a 4D Minkowski background).
It is interesting to recover from equations (\ref{HOPhiDGP}-\ref{HOdrDGP}) the results obtained in \cite{DGP}, and reminded in section \ref{DGPmodel}, for the Newtonian potential between two non relativistic sources. Consider indeed
a non relativistic source
which only non vanishing component of energy momentum tensor is  $\delta \rho_\MM$. Neglecting the time derivatives of the perturbations, equations (\ref{HOPhiDGP}) and (\ref{HOPsiDGP}) read 
\beq
\Phi&=& \frac{4}{3} \varphi_\bb, \label{Phiom}\\
\Psi &=& \frac{2}{3} \varphi_\bb, \label{Psiom}
\eeq
where we have defined $\varphi$ by $4 \varphi \equiv \Delta \Omega $.  
The master equation (\ref{FQ1}) together with equation (\ref{HOdrDGP}) can be recast in the unique equation
\beq 
\label{omeq}
 \frac{1}{\kcd} \left(\Delta \varphi + \varphi^{\prime \prime} \right) +  \frac{1}{\kqd}  \delta(y)  \Delta \varphi = \frac{1}{2} \delta(y) \delta \rho_\MM.  
\eeq
The above equation is the same as (\ref{GREENSCA})  and was solved in ref. \cite{DGP}. 
Considering a localized source, 
then $\varphi_\bb$,
as given by equation (\ref{omeq}), was found to vary for distances to the source
much smaller than $r_c$ as a usual four dimensional potential verifying (at small distances)
the standard Poisson equation (\ref{Poisson}) 
\beq  \label{POISS}
\Delta {\varphi}_\bb  = \frac{\kqd}{2} \delta \rho_\MM, 
\eeq
while at larger distances it was found to vary as a 5D potential  (see \cite{DGP}). 
One then recovers from the above discussion the two-folded manifestation of the vDVZ discontinuity: keeping $r_c$ much larger that the distance to the source, one sees first from equations (\ref{Phiom}) and (\ref{POISS}) that the gravitational potential 
on the brane $\Phi$ is renormalized by a factor $4/3$ with respect to the usual one. This 
has been discussed below equation (\ref{AMPLIT}).
Secondly, one has $\Psi = \frac{1}{2} \Phi$, which can also be derived from the  amplitude (\ref{AMPLITDGP}) and be seen to match analysis of the Schwarzschild solution 
in the DGP model \cite{DGP,Deffayet:2002uk,Gruzinov:2001hp,Giannakis:2002jg,Porrati:2002cp,Tanaka:2003zb}.

\subsection{General cosmological background: differential problem and wellposedness} \label{GCB}
When dealing with a more general cosmological background, the problem we are interested in 
becomes the following: given specified initial conditions (that one has to provide both on the brane and in the bulk), say at a given cosmological time $t_0$,  what are the perturbations as seen by a brane observer at a later cosmological time $t \geq t_0$? To answer this question for a given mode $\vec{k}$ requires solving the 2D PDE (\ref{MASTER2D}) with initial (Cauchy or also possibly characteristic) data provided along some initial curve in the 2D $(y,t)$ plane, and, given the hyperbolic nature of the problem, some boundary condition for $\Omega$ along the brane. In the following we will denote by $C_{(I)}$ the initial curve, and by $C_\bb$ the curve representing the brane trajectory in the $(y,t)$ plane. Those two curves are bounding a domain ${\cal D}$ where we need to solve the master equation (\ref{FQ1}). $C_{(I)}$ and $C_\bb$ intersect on the brane at some point $O$ which represents in the $(y,t)$ plane a space-like 3D initial hypersurface of cosmic time $t_0$.

In general there is not a clear way to give a boundary condition on the brane, since this requires solving the equations of motion for the perturbations of brane localized fields, which in turn are intricated with the metric perturbations. However in the simplest case of matter with vanishing anisotropic stress (for which (\ref{PHI1}-\ref{DP1}) hold), and simple equation of state, one can obtain easily such a boundary condition \cite{Kodama:2000fa,Deffayet:2002fn}. Indeed, for the case of adiabatic perturbations, 
substituting the expressions (\ref{DP1}) for $\delta P_{(M)}$, and (\ref{DR1}) for $\delta \rho_{(M)}$
in the equation of state (\ref{adiab})
yields a boundary condition on the brane of the form \cite{Kodama:2000fa,Deffayet:2002fn,DEF}
\beq
0&=&  \left\{
 \frac{\Delta \Omega}{a^2}  \CC^P_{\Delta (0,0)} + H {\Omega}^{\cdot}  \CC^{P}_{(1,0)} + \frac{\Delta {\Omega}^{\cdot}}{H a^2} \CC^{P}_{\Delta (1,0)}
+  {\Omega}^{\cdot \cdot}  \CC^P_{(2,0)}
+  \frac{\Delta {\Omega}^{\cdot \cdot}}{H^2 a^2}  \CC^P_{\Delta (2,0)}  \nonumber \right.\\&& \nonumber
 \left. +  \frac{\Omega^{\cdot \cdot \cdot}}{H} \CC^P_ {(3,0)}  + \frac{\Omega^{\cdot \cdot \cdot \cdot}}{H^2} \CC^P_ {(4,0)}
+  H \Omega^\prime \CC^P_{(0,1)} +  \Omega^{\cdot \prime}   \CC^P_ {(1,1)} +  \frac{\Omega^{\cdot \cdot \prime}}{H}   \CC^P_ {(2,1)}
\right\}_\bb
\\&&
- 3 c_S^2
 \left\{
 \frac{\Delta \Omega}{a^2}  \CC^\rho_{\Delta (0,0)} +  \frac{\Delta^2 \Omega}{H^2 a^4}  \CC^\rho_{\Delta^2 (0,0)}
+  H {\Omega}^{\cdot} \CC^\rho_{(1,0)} +   {\Omega}^{\cdot \cdot}  \CC^\rho_{(2,0)} 
+  \frac{\Delta {\Omega}^{\cdot \cdot}}{H^2 a^2}  \CC^\rho_{\Delta (2,0)}  \nonumber \right.\\&&
 \left.+ \frac{\Omega^{\cdot \cdot \cdot}}{H} \CC^\rho_ {(3,0)} + 
H \Omega^\prime \CC^\rho_{(0,1)} +  \frac{\Delta \Omega^\prime}{H a^2}  \CC^\rho_{\Delta (0,1)} +  \Omega^{\cdot \prime}   \CC^\rho_ {(1,1)}
\right\}_\bb.
\eeq
The latter can be recast in the form 
\beq \label{boundDGP}
0 = \sum_{r=0}^{r=4} \beta_{(r,0)} \partial^r_t \Omega 
+ \sum_{r=0}^{r=4} \beta_{(r,1)} \partial^r_t \Omega^\prime,
\eeq
where the $\beta$ are 
time dependent coefficients known from the background cosmological solution, and can be expressed as linear combinations of the coefficients ${\cal C}$. 
A boundary condition of identical form is found in the case 
where the only matter on the brane is a scalar field, substituting the expressions (\ref{DP1}) for $\delta P_{(M)}$, 
(\ref{DQ1}) for $\delta q_{(M)}$, and (\ref{DR1}), into equation (\ref{STATSCA}).
In the following we will restrict ourselves to the here-above mentioned  cases where a 
boundary condition of the form (\ref{boundDGP}) holds.

Equations (\ref{FQ1}) and (\ref{boundDGP}) are then all what is needed to solve for the evolution of DGP brane world cosmological perturbations once initial conditions are supplied in the bulk. Those two equations play for the brane world cosmological perturbations of the model at hand, an equivalent r\^ole to the one played by equation (\ref{EVOL}) for 4D adiabatic cosmological perturbations
of a perfect fluid. One might worry about the non standard form of the boundary condition (\ref{boundDGP}) which involves derivatives of the master variable along the brane (such a boundary condition has been called {\it non local} \cite{Kodama:2000fa}) . However, one can recast the differential problem defined by equations (\ref{FQ1}) and (\ref{boundDGP}) in a standard form \cite{DEF}. 
This is done as follows. First we introduce the characteristic coordinates $X$ and $Y$ of the differential operator ${\cal D}_\Delta$, such that the master equation is rewritten as 
\beq
\partial_{X}\partial_{Y} \Omega &=& \frac{3 \partial_Y \Omega}{2 X} + \frac{\Delta \Omega}{4 X^2} \label{MASTERXY1} \\ \label{MASTERXY}
 &\equiv& V(X) \partial_Y \Omega + W_\Delta(X) \Omega,
\eeq where the coefficients $V$ and $W_\Delta$ are functions of $X$ (and also of the wave number $\vec{k}$, as far as $W_\Delta$ is concerned) defined by the first line \footnote{Note that the differential operator (\ref{MASTERXY1}) is singular in $X=0$, which corresponds to a coordinate singularity (see \cite{Deffayet:2002fn}) in the characteristic $(X,Y,x^i)$ coordinatization of 5D Minkowski space-time. For the rest of the discussion one can (but does not necessarily have to) always consider a domain ${\cal D}$ which does not intersect the $X=0$ line.}.
One then defines the 15 unknowns 
$u_{(r,s)}$, 
with $0 \leq r \leq 4 $, $0 \leq  s \leq 4$, and $0\leq r+s \leq 4$ by 
\beq
u_{(r,s)}= \partial^r_X \partial^s_Y \Omega, 
\eeq
and recast the differential equation (\ref{MASTERXY}) in the following linear system 
\beq \label{equa1}
\partial_X u_{(r,s)} &=& u_{(r+1,s)} ,\\&& 
 \quad {\rm for} \quad r+s < 4, \nonumber \\ \label{equa2}
\partial_X u_{(r,s)} &=& \partial^{r}_X \partial^{s-1}_Y \left(V u_{(0,1)} + W_\Delta  u_{(0,0)} \right) ,\label{deux}  \\
&&\quad {\rm for} \quad r+s = 4 \quad {\rm and} \quad s \geq 1, \quad {\rm and} \nonumber\\ \label{equa3}
 \partial_Y u_{(4,0)} &=& \partial^{3}_X  \left( V u_{(0,1)} + W_\Delta  u_{(0,0)} \right). \label{trois}
\eeq
The boundary condition (\ref{boundDGP}) then takes the form 
\beq \label {BOUNTY}
u_{(4,0)} = \sum_{(r,s), r+s\leq 4} \alpha_{(r,s)} u_{(r,s)},
\eeq
where $\alpha_{(4,0)}$ vanishes by definition and the others $\alpha_{(r,s)}$ are known functions of the parameter $t$ along the brane (the cosmic time) that can easily be obtained from the expression (\ref{boundDGP}). 
This rewriting of eq. (\ref{boundDGP}) is possible as long 
as the coefficient of the highest derivative of $\Omega$ appearing in (\ref{boundDGP}), namely $\Omega^{\cdot \cdot \cdot \cdot}$ for the cases considered here, does not vanish. The latter coefficient is given by 
\beq
\frac{2 + \Upsilon}{2 a_\bb \kappa_{(4)}^2 \left(\Upsilon + \frac{2 H^2 +\dot{H}}{H^2}\right)}.
\eeq
It does not vanish for a sufficiently large class of cosmological backgrounds.
The system (\ref{equa1}-\ref{trois}) is a linear first order system written in its so called characteristic normal form, with a boundary condition (\ref{BOUNTY}) of standard form on a non characteristic curve $C_\bb$ (the brane is non characteristic, since it follows a time-like trajectory).   
Standard theorems on linear differential systems then insure that the differential problem associated with the system (\ref{equa1}-\ref{equa3}) and the boundary condition (\ref{BOUNTY}) is well posed\footnote{Note in particular that there is one boundary condition (\ref{BOUNTY}) for one in-going characteristic $X=constant$ curve in the domain ${\cal D}$ of interest, as it should (see e.g. \cite{courant}).}
provided one gives initial data of Cauchy type on the initial curve $C_{(I)}$, that is to say the values there of all the variables $u_{(r,s)}$, with $r+s \leq 4$. 
Under this condition, the problem has a unique solution in the domain ${\cal D}$, and this solution depends continuously on any parameter  which initial and boundary data may depend on.

However, because $\Omega$ obeys in the bulk an hyperbolic PDE of order two, one cannot freely specify on $C_{(I)}$ all the $u_{(r,s)}$ in an independent way without encountering inconsistencies. E.g., considering first the case where $C_{(I)}$ is non characteristic in the vicinity of $O$ (and remains so all along the segment of the initial curve where we want to specify initial data), then it is enough to provide on the initial curve the value of $\Omega$ and of its normal  
derivative with respect to $C_{(I)}$ in order to know there the initial values of all the 
$u_{(r,s)}$ with $r+s \leq 4$. This follows directly from the fact that the initial curve is a non characteristic curve for the PDE (\ref{FQ1}).  On the other hand, if the initial curve is a characteristic in the vicinity of $O$, then, in order to know the initial values of all the $u_{(r,s)}$ (with $r+s \leq 4$), 
it is 
 enough to specify on $C_{(I)}$ the value of $\Omega$ as well as
to specify in $O$, and only there, the values of the derivatives of $\Omega$ normal to $C_{(I)}$ of order lower or equal to four (see \cite{DEF} for more details).
Note that once the values of the $u_{(r,s)}$ are known on the initial curve $C_{(I)}$, and thus also at O, then all the initial values (i.e. at the initial time $t_0$) of the brane variables $\Phi$, $\Psi$, $\delta \rho_\MM$, $\delta q_\MM$, $\delta P_\MM$ are known through equations (\ref{PHI1}-\ref{DP1}). So that if one wishes to study a specific case where those initial values are specified, one should take care to choose the initial values of the $u_{(r,s)}$ (or rather, as we just saw, those of $\Omega$ and its derivatives) accordingly. 
Moreover, as it will turn out to matter for the discussion in the next subsection, one should also demand that, in all case, the chosen initial values of the $u_{(r,s)}$
 are compatible with the boundary condition (\ref{BOUNTY}). This is a standard requirement for a hyperbolic problem with a boundary.

\subsection{Disappearance of the vDVZ discontinuity on FLRW background
with non vanishing scalar curvature
} \label{vDVZ2}
We are now in a position to discuss the vDVZ discontinuity on a FLRW background. We 
want to study the behavior of cosmological perturbations of the DGP model where $r_c$
 is send to infinity with respect to a background cosmological length scale, that we will take to be the Hubble radius $H^{-1}$. In this
 limit: i/ we want to see whether initial data for the brane quantities $\Phi$, $\Psi$, $\delta \rho_\MM$, $\delta q_\MM$, $\delta P_\MM$,  
can be specified arbitrarily close to the initial data for the corresponding quantities of usual 4D cosmological perturbations (or to say it in an other way, if the initial constraints of 4D GR can be fulfilled), and, ii/ we want to know if the evolution of those initial data differ, or not, from the one given by the theory of standard 4D cosmological perturbation as recalled in section \ref{4Dpert}. These two questions are the ones an observer living on the brane would like to answer in order to compare in a cosmological context the predictions of the linearized DGP model to the ones of linearized 4D GR.  

Thus, the limit we will subsequently consider is 
$H r_c  \rightarrow \infty$, that is to say $\Upsilon \rightarrow 0$, 
keeping $H$, $\kqd$, as well as the other 4D background quantities $\rho_\MM$ and $P_\MM$, finite. This simply because we want to compare cosmological perturbations of the DGP model with the ones of standard GR in the same cosmological background. This limit will be denoted simply by $\Upsilon \rightarrow 0$. 

Let us first discuss the case of the background space-time, in a way appropriate for the discussion of cosmological perturbations. 
As we already explained below equation (\ref{Upsi}), in the $\Upsilon \rightarrow 0$ limit, the brane Friedmann equations (\ref{fried1}-\ref{fried2}) go to the standard Friedmann equations
\beq
\dot{\rho}_\MM &=& - 3 H (P_\MM + \rho_\MM), \label{stanfried1}\\
3 H^2 &=& \kqd \rho_\MM \label{stanfried2},
\eeq
obtained simply by making $\Upsilon = 0$ in the system (\ref{fried1}-\ref{fried2}).
For a specific equation of state for the background matter $\rho_\MM$ and $P_\MM$, the systems (\ref{stanfried1}-\ref{stanfried2}) and (\ref{fried1}-\ref{fried2}) are both second order differential system in the scale factor $a_\bb$. 
This means in particular that one can freely choose the initial values of $a_\bb$ and $\dot{a}_\bb$, at the initial time $t_0$.
The brane Friedmann equations define subsequently a brane trajectory in the characteristic coordinates $X$ and $Y$. The brane curve $C_\bb$ is indeed given by \cite{Deffayet:2002fn}
\beq
X_\bb &=& a_\bb, \\
\dot{Y}_\bb &=& \dot{a}^{-1}_\bb.  
\eeq
We denote by $C_\bb^{\Upsilon\neq0}$ any brane curve, which is given by the solution of the system (\ref{fried1}-\ref{fried2}) with 
a non vanishing
 $\Upsilon$, and goes across some fixed point O in the $(X,Y)$ plane, at which we fix accordingly the initial value 
for $a_\bb$ and $\dot{a}_\bb$ at the initial time $t_0$. We also define $C_\bb^{\Upsilon = 0}$ as the curve 
intersecting the previous ones in O, defined by the same initial values for $a_\bb$ and $\dot{a}_\bb$ in O, but the system (\ref{stanfried1}-\ref{stanfried2}). One sees that the limiting curve $C_\bb^{\Upsilon \rightarrow 0}$ is exactly $C_\bb^{\Upsilon = 0}$. This is  only a geometric formulation of the fact that the early time DGP cosmology is identical to standard cosmology, as noted in \cite{Deffayet:2001uy}, and that there is no discontinuity in the background cosmological solution. 
Note moreover that 
in the same limit as above, the background geometry stays perfectly regular (it is always a slice of 5D Minkowski space-time) and the $(y,t)$ coordinate system does not break down in the vicinity of the brane, as can be seen from equations (\ref{a}) and (\ref{n}). 
The conclusion of all this, is that in order to compare the outcome 
of cosmological perturbations in the DGP model and in standard 4D GR on a given FLRW background, it is enough to choose for the background equation of motion 
(\ref{stanfried1}-\ref{stanfried2}) and (\ref{fried1}-\ref{fried2}), the same initial values for $a_\bb$ and $\dot{a}_\bb$, and the same matter equation of state for both model. This is because the resulting trajectory 
$C_\bb^{\Upsilon\neq0}$ can be made arbitrarily close to the {\it purely 4D} trajectory $C_\bb^{\Upsilon = 0}$ by taking a sufficiently small $\Upsilon$. 
The difference between the two curves will then only influence the behavior of cosmological perturbations at the second non trivial order in 
$\Upsilon$, which will be enough for our discussion. 
An alternative approach, which might be considered less physical since it requires an $\Upsilon$-dependent equation of state,  is to choose exactly the {\it same} curves 
$C_\bb^{\Upsilon\neq0}$ and $C_\bb^{\Upsilon = 0}$ by taking 
a different equation of state for the background matter  in the $\Upsilon \neq 0$ case and the  $\Upsilon = 0$ case. 
This leads to a slightly simpler discussion since the background brane space-time are identical in the 
$\Upsilon \neq 0$ case and in the  $\Upsilon = 0$ case.
We will only consider in this subsection the first case: namely a given fixed  equation of state (or fixed matter equations of motion) for the background matter on the brane, with the understanding that this will result in an $\Upsilon$-dependence of the domain ${\cal D}$ over which we study the master equation. However we stress again that the two approaches are equivalent as far as the conclusions are concerned. We will moreover assume in this subsection that the limiting trajectory, $C^{\Upsilon =0}_\bb$, is such that the Ricci scalar, $R^{(B)}_\bb$, of the induced metric on the brane does not vanish on the cosmic time interval $[t_0,t_{\rm max}]$ over which we want to follow the evolution of the cosmological perturbations. This can always be done choosing initial conditions and the matter equation of state, and will also imply that the Ricci scalar of the background metric on the brane, for $\Upsilon \neq 0$, does not vanish on the time interval of interest, for small enough $\Upsilon$.

Let us now turn to the behavior of the perturbations themselves. The associated differential problem is entirely specified by the master equation (\ref{FQ1}), the boundary condition (\ref{boundDGP}), and a domain ${\cal D}$ bounded by the brane $C_\bb$ and the initial curve $C_{(I)}$ on which we give initial data. Among those different elements, we have just discussed the behavior of  $C_\bb$ in the $\Upsilon \rightarrow 0$ limit. As far as $C_{(I)}$ is concerned, we  pick a fixed initial curve in the bulk, independently of $\Upsilon$ (which we can always do). The curve $C_{(I)}$, which intersects all the $C^\Upsilon_\bb$ curves in O,  is chosen so that it is non characteristic (a similar discussion with identical conclusions can easily be carried out when the initial curve is characteristic). 
Note that, when $\Upsilon$ does not vanish, the domain ${\cal D}$ can be made arbitrarily close to the limiting domain bounded by the chosen initial curve and the limiting curve  
$C_\bb^{\Upsilon \rightarrow 0}$. The master equation does not depend on $\Upsilon$ (see e.g. equation (\ref{MASTERXY1})), so the only $\Upsilon$-depending pieces that we have not discussed are the choice of initial data on the initial curve $C_{(I)}$ and the boundary condition (\ref{boundDGP}). 
Let us now turn to these last two issues.

We first note one crucial result: assuming that $R^{(B)}_\bb$ does not vanish on $[t_0,t_{\rm max}]$, it is easily seen, looking at the expressions of the coefficients ${\cal C}$ given in the appendix \ref{expressC}, that all of those coefficients have a well behaved and bounded limit\footnote{Provided one considers a background space-time where $H$ and its time-derivatives are bounded on the time interval of interest. Note that one would have encountered divergences going as $1/\Upsilon$ at any point where $R^{(B)}_\bb$ does not vanish. This will be further discussed in subsection \ref{vDVZ4}.} when $\Upsilon \rightarrow 0$. This limit is continuous and the limiting coefficients, as function of $H$ and its time derivatives, are simply given by the expressions of appendix \ref{expressC} taken in $\Upsilon = 0$. A first consequence of this is that the boundary condition (\ref{boundDGP}) has a well defined limit when $\Upsilon \rightarrow 0$.
 
We turn then to the choice of initial conditions. For future convenience, we introduce first a new coordinate system $\tau(t,y)$ and $z(t,y)$ such that the curve  $C_{(I)}$ is defined by a constant $z$, and that the $\tau =$ constant curves are normal to $C_{(I)}$. For a fixed $\Upsilon$, we can choose arbitrarily on $C_{(I)}$ the values of $\Omega$ and of its normal derivative $\partial_z \Omega$ w.r.t. $C_{(I)}$. The only constraint these values have to obey comes from requiring their compatibility with the boundary condition (\ref{boundDGP}) in O, which depends on $\Upsilon$.
In addition, having in mind a comparison with cosmological perturbations 
 of 4D GR, we wish to fix at the initial time $t_0$ on the brane, that is to say at O, the values  $\Phi_{t_0}$ and $\dot{\Phi}_{ t_0}$ (of $\Phi$ and $\dot{\Phi}$ respectively). We thus impose three constraints on the initial conditions at O, which in turn impose constraints on the value of $\Omega$ and of its derivatives at O, through equations (\ref{boundDGP}), (\ref{PHI1}), and the derivative of (\ref{PHI1}) with respect to $t$. That these constraints can be simultaneously imposed at O, simply by choosing the values of  $\Omega$ and  $\partial_z \Omega$ 
on $C_{(I)}$ is easy to see. One way to proceed goes as follows. 
Using the fact the initial curve is assumed non characteristic, one can express the derivatives $\Omega^{\cdot \cdot}$, $\Omega^{\cdot \cdot \cdot}$ and $\Omega^{\cdot \cdot \cdot \cdot}$ at O (that we denote respectively as 
 $\Omega^{\cdot \cdot}_{t_0}$, $\Omega^{\cdot \cdot \cdot}_{t_0}$ and $\Omega^{\cdot \cdot \cdot \cdot}_{t_0}$) as 
\beq \label{OddtO}
\Omega^{\cdot \cdot}_{t_0} &=& \frac{2 \dot{z} z^\prime \left(\dot{z}\tau^\prime - z^\prime \dot{\tau}\right)}{\dot{z}^2 - {z}^{\prime 2}}  \partial_{z}\partial_{\tau}\Omega + {\cal O}(2) \\ \label{OdddtO}
\Omega^{\cdot \cdot \cdot}_{t_0} &=& 
 \frac{\dot{z}(\dot{z}^2 + 3 z^{\prime 2}) \left(\dot{z}\tau^\prime - z^\prime \dot{\tau}\right)^2}{\left(\dot{z}^2 - {z}^{\prime 2}\right)^2}  \partial_{z}\partial^2_{\tau}\Omega + {\cal O}(3)\\\label{OddddtO}
 \Omega^{\cdot \cdot \cdot \cdot}_{t_0} &=& 
 \frac{4 \dot{z} z^\prime (\dot{z}^2 + z^{\prime 2}) \left(\dot{z}\tau^\prime - z^\prime \dot{\tau}\right)^3}{\left(\dot{z}^2 - {z}^{\prime 2}\right)^3}  \partial_{z}\partial^3_{\tau}\Omega + {\cal O}(4)
\eeq
where ${\cal O}(n)$ denotes terms which are linear combinations of derivatives of $\Omega$,  $\partial^n_\tau \Omega$ or $\partial^s_z \partial^r_\tau  \Omega$ with $r+s < n$ and $s \leq 1$. \label{VDVZPAR11}
We then choose arbitrary values for $\Omega$ all along $C_{(I)}$. This means that all the derivatives of $\Omega$ of the form 
$\partial^r_\tau  \Omega$ at known in O. In addition we pick some arbitrary value for $\partial_z \Omega$ at O. Then, inserting the expression (\ref{OddtO}) in equation (\ref{PHI1}) we determine the value at O of $\partial_z \partial_\tau \Omega$ as a function of the chosen value for $\Phi_{t_0}$. In a similar way we can determine the value at O of $\partial_z \partial^2_\tau \Omega$ from the chosen value for  $\dot{\Phi}_{t_0}$ and the derivative w.r.t. t of equation (\ref{PHI1}). Last the value of $\partial_z \partial^3_\tau \Omega$ at O is determined from the boundary condition ({\ref{boundDGP}) at O. At this point the values at O of all the derivative  $\partial_z \partial^r_\tau \Omega$, with $r\leq 3$ are known, and we complete the initial condition by picking any function 
 $\partial_z \Omega$ along $C_{(I)}$ compatible with these values at O.  
Note that the above construction is possible because: i/ the highest derivatives of $\Omega$ entering in the expression of $\Phi$ and $\dot{\Phi}$, obtained from equation (\ref{PHI1}), and in the boundary condition (\ref{boundDGP}) are respectively $\Omega^{\cdot \cdot}$, $\Omega^{\cdot \cdot \cdot}$, and $\Omega^{\cdot \cdot \cdot \cdot}$, and, ii/ the coefficients multiplying $\partial_z \partial_\tau \Omega$, $\partial_z \partial^2_\tau \Omega$, $\partial_z \partial^3_\tau \Omega$ in the above expression (\ref{OddtO}), (\ref{OdddtO}), (\ref{OddddtO}), do not vanish, nor diverge, because of the invertible character of the coordinate change $z(t,y), \tau(t,y)$, and the non characteristic nature of $C_{(I)}$ (which insure that $\dot{z}^2 - z^{\prime 2}$ does not vanish). 
More generally we call $\Omega^\Upsilon_{(I)}$ and $\partial_z \Omega^\Upsilon_{(I)}$ any set of initial condition on $C_{(I)}$ which are compatible with the three constraints mentioned above, for a fixed value of $\Upsilon$. We will assume that the functions $\Omega^\Upsilon_{(I)}$ and $\partial_z \Omega^\Upsilon_{(I)}$ have  continuous limits when we let $\Upsilon$ go to zero. We call ${\cal S}$ a set of initial conditions (for $\Upsilon \in$  some interval $]0,
\Upsilon_{\rm max}]$) having this property. Again inspection of the coefficients ${\cal C}$ and of equations  (\ref{OddtO}), (\ref{OdddtO}) and (\ref{OddddtO}) show that such a set is not empty. 
To summarize, a set of initial conditions 
\beq \label{SETIN}
{\cal S}= \left\{\left(\Omega^\Upsilon_{(I)},\partial_z \Omega^\Upsilon_{(I)}\right),\Upsilon \in  ]0,\Upsilon_{\rm max}]\right\}
\eeq
 has the property that its elements $ 
\left(\Omega^\Upsilon_{(I)},\partial_z \Omega^\Upsilon_{(I)}\right)$ fulfill the boundary condition (\ref{boundDGP}) in O, give fixed values $\Phi_{t_0}$ and $\dot{\Phi}_{t_0}$ for $\Phi$ and $\dot{\Phi}$ at O, and go to a well defined limiting initial condition when $\Upsilon$ is send to zero. 
We call  $\Omega^\Upsilon_{\cal S}$ the solution over ${\cal D}$ to the master equation (\ref{FQ1}) with the initial data 
$\left(\Omega^\Upsilon_{(I)},\partial_z \Omega^\Upsilon_{(I)}\right)$ chosen in a given set ${\cal S}$ and obeying the boundary condition (\ref{boundDGP}). 
The here-above noticed fact that the boundary condition (\ref{boundDGP}) has a well behaved limit when $\Upsilon \rightarrow 0$, 
 as well as the fact that the differential problem we are looking at is well posed \cite{DEF}, shows that the solution $\Omega^\Upsilon_{\cal S}$ has a well defined limit when $\Upsilon \rightarrow 0$, we call this limit  $\Omega^{\Upsilon\rightarrow 0}_{\cal S}$. 
Once one knows this limit, one can compute from equations (\ref{PHI1}), (\ref{PSI1}), (\ref{DR1}), (\ref{DQ1}) and (\ref{DP1}), the limiting value (as $\Upsilon \rightarrow 0$) at all time $t \in [t_0, t_{\rm max}]$ of the brane expressions $\Phi$, 
$\Psi$, $\delta \rho_\MM$, $\delta q_\MM$, $\delta P_\MM$. The question of the vDVZ discontinuity then translates into asking how these limiting values, noted $\Phi^{\Upsilon \rightarrow 0}$, 
$\Psi^{\Upsilon \rightarrow 0}$, $\delta \rho_\MM^{\Upsilon \rightarrow 0}$, $\delta q_\MM^{\Upsilon \rightarrow 0}$, $\delta P_\MM^{\Upsilon \rightarrow 0}$,  compare with the corresponding quantities of standard 4D GR, than is to say the solutions of equations (\ref{EQM6_4D}-\ref{EQM8_4D}) with the same initial conditions $\Phi_{t_0}$ and $\dot{\Phi_{t_0}}$.

To answer this question, we first  
  define new quantities, that we will denote respectively  
 $\Phi^0$, 
$\Psi^0$, $\delta \rho_\MM^0$, $\delta q_\MM^0$, $\delta P_\MM^0$, by  taking the limiting values in $\Upsilon \rightarrow 0$ of the right hand side of equations (\ref{PHI1}-\ref{DP1}). E.g.  $\Phi^0$ is defined by 
\beq \label{PHI2}
\Phi^0 &=& \frac{1}{6 a_\bb} \left\{ \frac{\Delta \Omega}{a^2}  \CC^{\Phi}_{\Delta (0,0)}
+ H {\Omega}^\cdot  \CC^{\Phi}_{(1,0)} + {\Omega}^{\cdot \cdot}   \CC^{\Phi}_{(2,0)} + 
 H \Omega^\prime \CC^{\Phi}_{(0,1)} \right\}_{\bb, \Upsilon=0},
\eeq
where the index $_{\Upsilon=0}$ means that all the  coefficients ${\cal C}$ are taken in $\Upsilon =0$, but also that the 
quantities depending on the background brane trajectory (and thus also on $\Upsilon$, as we saw above) are given by the limiting $C_\bb^{\Upsilon =0}$ trajectory. This means e.g. that the $H$ entering into (\ref{PHI2}) is obtained by definition by solving the standard Friedmann equations (\ref{stanfried1}-\ref{stanfried2}). Note further that the function $\Omega$ entering into the above definition of $\Phi^0$, $\Psi^0$, $\delta \rho_\MM^0$, $\delta q_\MM^0$, and $\delta P_\MM^0$ is so far left arbitrary. 
Now comes the remarkable fact: 
one can verify after some straightforward, but tedious, algebra that the expressions $\Phi^0$, 
$\Psi^0$, $\delta \rho_\MM^0$, $\delta q_\MM^0$, $\delta P_\MM^0$
verify identically  the standard 4D perturbed Einstein's equations (\ref{EQM6_4D}-\ref{EQM8_4D}) with $\delta \pi_\MM =0$, for arbitrary $\Omega$ and $\Omega^\prime$ (in particular $\Phi^0$ equals $\Psi^0$). 
This answers the question, since it shows that for any set if initial condition ${\cal S}$ (and thus any limiting solution $\Omega^{\Upsilon \rightarrow 0}_{\cal S}$) the limiting expressions $\Phi^{\Upsilon \rightarrow 0}$, 
$\Psi^{\Upsilon \rightarrow 0}$, $\delta \rho_\MM^{\Upsilon \rightarrow 0}$, $\delta q_\MM^{\Upsilon \rightarrow 0}$, $\delta P_\MM^{\Upsilon \rightarrow 0}$
are exactly equal to the corresponding solutions of cosmological perturbations computed from 4D GR, with the same initial conditions.
Thus the latter can not be distinguished from the former by a brane observer.  
In this sense, it shows that there is no vDVZ discontinuity on a FLRW space-time with non vanishing Ricci scalar curvature (with the restriction that our proof is only valid for matter with vanishing anisotropic stress and equation of state 
 leading to a boundary condition of the form (\ref{boundDGP})).

An other way to look at this result is to say that in the $\Upsilon \rightarrow 0$ limit, the knowledge at O of $\Phi$ and $\dot{\Phi}$ and of the brane matter equation of state are enough to compute on $C_\bb$ the values of 
$\Phi$, 
$\Psi$, $\delta \rho_\MM$, $\delta q_\MM$, $\delta P_\MM$
 at all time $t \in [t_0, t_{\rm max}]$. So that the brane decouples from the bulk in this limit. And a Dirichlet boundary condition for the operator (\ref{FQ1}) is then obtained extracting  from equations (\ref{PHI1}), (\ref{PSI1}), (\ref{DR1}), (\ref{DQ1}) and (\ref{DP1}) (and (\ref{FQ1})) the expression of 
$\Omega$ as a function of $\Phi$, 
$\Psi$, $\delta \rho_\MM$, $\delta q_\MM$, $\delta P_\MM$ (this is done in Ref. \cite{Deffayet:2002fn}, the corresponding equation is obtained from equations (\ref{Omat}) and (\ref{Er}) of this paper).

\subsection{vDVZ discontinuity on Minkowski background, as seen in time-dependent perturbations}
It is instructive to contrast the case we have just discussed, with the case of  time dependent perturbations over a Minkowski background\footnote{Time dependent aspects of the vDVZ discontinuity of the Pauli-Fierz theory have been studied in ref. \cite{VanNieuwenhuizen:1973qf}.}. 
First, we consider the case of a constant $H$ background (so that the background is either de Sitter or Minkowski). It is easy to obtain from appendix \ref{expressC}, the expressions (\ref{PHI1}-\ref{DP1}) corresponding to that case, by dropping all the time derivatives of $H$ in the coefficients ${\cal C}$. Doing that, the coefficients ${\cal C}$ are found to be polynomials of $\Upsilon$ and $f_\Upsilon$; e.g., for a constant $H$ background, ${\cal C}^q_{\Delta(0,0)}$ is given by 
$f^2_\Upsilon (24 + 24 \Upsilon + 6 \Upsilon^2)$. 
To obtain the expressions (\ref{HOPhiDGP}-\ref{HOdPDGP}) one then sets $H$ to zero in the constant-$H$ expressions. The essence of the vDVZ discontinuity, in the way discussed here, is  that the limits $H \rightarrow 0$, and $r_c \rightarrow \infty$, do not commute. Indeed if one sets $H$ to a constant value, and then let $\Upsilon$ go to zero, one obtains for example the following expression for $\Phi^0$,
\beq \label{PHIDESIT}
\frac{1}{4 a_\bb}\left\{ \frac{\Delta \Omega}{a^2} + 3 H \Omega^{\cdot} - \Omega^{\cdot \cdot} + 2 \eta H \Omega^\prime \right\}_\bb,
\eeq 
which is equal to the expression of $\Psi^0$, as it should be from our previous discussion, but stays different from (\ref{HOPhiDGP}) and (\ref{HOPsiDGP}) in the $H \rightarrow 0$ limit\footnote{Equation (\ref{PHIDESIT}) also illustrates the fact that, on a cosmological background, the perturbation theory depends explicitly on the phase $\eta =\pm 1$ of the background, this even for distance scales unrelated to $r_c$, as first noticed in reference \cite{SOLAR}.}.

We would like now to ask the same question as in subsection \ref{VDVZPAR11}, but for time dependent perturbations over a Minkowski brane. First we recall that if we set $\delta \pi_\MM$ and $\delta P_\MM$ to zero in equations (\ref{EQM6_4D}-\ref{EQM8_4D}) (with $\dot{a}=0$), or equivalently $c_S$
to zero in equation (\ref{EVOL}), one finds that $\Phi$ equals $\Psi$ at all time and is given by a affine function of time. If one further demands that the brane is empty at the initial time $t_0$ (that is to say 
 that $\delta \rho_\MM$ and  $\delta q_\MM$ vanish at initial time), then the initial values of $\Phi$ and $\dot{\Phi}$ have to vanish thanks to the initial constraints (\ref{EQM6_4D}) and (\ref{EQM4_4D}), meaning further that $\Phi$ vanish at all later time. As is well known, there are no scalar gravitational waves in 4D GR.
Let us now discuss the case of the DGP model, with a Minkowski brane, the same initial conditions, and an empty brane. 
The master equation (\ref{FQ1}) takes the form
of a usual 5D d'Alembertian equation
\beq \label{FQ1MINK}
{\Omega}^{\cdot \cdot} - \Omega^{\prime \prime } + \vec{k}^2 \Omega = 0,
\eeq
a solution of which can be taken to be simply the ``zero mode'' 
\beq \label{solution}
\Omega = A \sin \left(\omega(t-t_0) + \omega_0\right),
\eeq
with $\omega^2 = \vec{k}^2$. 
The $\delta \rho_\MM$ defined from equation (\ref{HOdrDGP}) and this solution vanishes identically, as well as the $\delta q_\MM$ and $\delta P_\MM$, so that 
this solution verifies the boundary condition for an empty brane. On the other hand the initial values $\Phi_{t_0}$ and $\dot{\Phi}_{t_0}$ can be chosen arbitrarily by choosing the amplitude $A$ and the phase $\omega_0$ to verify
\beq 
A &=&  \frac{6}{\omega^3} \sqrt{\dot{\Phi}^2_{t_0} + \omega^2 \Phi^2_{t_0}}, \nonumber \\
\tan \omega_0 &=& \frac{\omega \Phi_{t_0}}{ \dot{\Phi}_{t_0}}, \label{BOUNDPHIPHID}
\eeq
as can be seen from equation (\ref{HOPhiDGP}). 
There are thus  propagating scalar gravitational waves on the brane in the linearized DGP model. Those waves verify $\Psi= -\Phi$, as can be seen from equation (\ref{HOPsiDGP}) and (\ref{solution}),
in contrast with 4D GR, where the vanishing of $\delta \pi_\MM$ implies the equality between $\Phi$ and $\Psi$. If one wants then to stick with the reasoning of section \ref{vDVZ2}, one can choose initial data on some fixed initial curve $C_{(I)}$ such that the solution (\ref{solution}) with the choices (\ref{BOUNDPHIPHID}) will be the unique solution corresponding to an empty brane, initial values $\Phi_{t_0}$ and $\dot{\Phi}_{t_0}$, and those initial data. The initial data, so defined on 
$C_{(I)}$, will obviously be independent of $r_c$, as is the solution (\ref{solution}), so that both will be unchanged in the $r_c \rightarrow \infty$ limit. This provides an example that the recovery of solutions of 4D GR cosmological perturbation theory shown to occur on a FLRW brane with non vanishing scalar curvature, does not happen for a Minkowski brane\footnote{It is quite simple to find other solutions showing the same, with $r_c$ dependent initial data.}. 
\label{vDVZ3}

\subsection{Case of a brane with non zero Hubble factor, but vanishing scalar curvature} \label{vDVZ4}
We now turn to the case of a background with a vanishing Ricci scalar $R_\bb^B$. Namely, we will assume that $R_\bb^B$ vanishes at all times in the time interval $[t_0, t_{max}]$. This can be achieved as discussed in subsection \ref{VDVZFRW1p14}. In this case, the reasoning done in section \ref{VDVZPAR11} is no longer valid because the coefficients ${\cal C}$ all diverge as $1/\Upsilon$, when  $\Upsilon$ is sent to zero. However, one can easily get rid of this divergence by considering the new variable $\tilde{\Omega}$ defined by 
\beq \label{DEFOMTILDE}
\Omega = \Upsilon \tilde{\Omega}.
\eeq 
$\tilde{\Omega}$ obeys a PDE in the bulk that can be obtained from (\ref{FQ1})
 and reads 
\beq 
\label{FQ1bis}
\left(\frac{1}{na^3}\tilde{\Omega}^\cdot\right)^\cdot 
+ 2 H \left(\frac{1}{na^3}\right)^\cdot \tilde{\Omega} 
+  4 H \left(\frac{1}{na^3}\right)\tilde{\Omega}^\cdot - 
\frac{n }{a^5}\Delta \tilde{\Omega} - \left(\frac{n
}{a^3}\tilde{\Omega}^\prime \right)^\prime=0,
\eeq
where $H$ means $\dot{a}_\bb/a_\bb$, as in the preceding. To obtain this expression, we have used the vanishing of $R^B_\bb$ which implies that 
\beq
\dot{\Upsilon} &=& \frac{2 \eta}{r_c} \nonumber,\\
&=& 2 H \Upsilon. 
\eeq
It might seem that the above PDE depends implicitly on $\Upsilon$ through $H$. This is however not the case, since $H$ is fully determined by solving the equation $2H^2 + \dot{H}=0 $, once the initial values of $a_\bb$ and $\dot{a}_\bb$ have been provided. This in turn also determines the brane trajectory in the characteristic coordinates $X$ and $Y$ of the differential operator associated with equation (\ref{FQ1bis}). Note that the latter coordinates are the same as those previously introduced, since they only depend on the derivatives of second order in the differential operator considered. Substituting $\Omega$ by $\Upsilon \tilde{\Omega}$ in equations (\ref{PHI1}-\ref{DP1}) one obtains expressions for $\Phi$, $\Psi$, $\delta \rho_\MM$, $\delta q_\MM$, and $\delta P_\MM$, as functions of $\tilde \Omega$ and its derivatives, with new coefficients $\tilde{\cal C}$, instead of ${\cal C}$. One can verify that those new coefficients have all now a well behaved limit when $\Upsilon$ is let to zero. 
In addition, one can show as in section \ref{VDVZPAR11} that the limiting expression of  $\Phi$, $\Psi$, $\delta \rho_\MM$, $\delta q_\MM$, and $\delta P_\MM$, defined as 
$\Phi^0$, $\Psi^0$, $\delta \rho^0_\MM$, $\delta q^0_\MM$, and $\delta P^0_\MM$, with $\tilde{\cal C}$ and $\tilde{\Omega}$ replacing respectively ${\cal C}$ and $\Omega$ verify identically the linearized 4D GR equations (\ref{EQM6_4D}-\ref{EQM8_4D}), whatever the function $\tilde{\Omega}$. This means that the rest of the discussion follows as in \ref{VDVZPAR11}. 

The case discussed here is however quite different from that discussed in subsection \ref{VDVZPAR11}. Indeed, the limiting values of $\tilde{\Omega}$  and its derivatives are bounded in the vicinity of the brane, which means that those of $\Omega$ and its derivatives vanish by virtue of the definition 
(\ref{DEFOMTILDE}). It implies  that the limiting values of the four 5D scalar gauge invariant variables (\ref{AGIV}-\ref{RGIV}) vanish; this because those gauge invariant variables are all expressed as linear combinations of $\Omega$, its derivatives, and the background metric \cite{Mukohyama:2000ui,Kodama:2000fa} (see equations (\ref{EQ1}-\ref{EQ20})). On the other hand the limiting value of the gauge invariant brane position $\tilde{\xi}$ is finite, as can be seen from equation (\ref{POSPHIPSI}). Thus, the limiting regime is describing a brane moving in an {\it unperturbed} bulk, that is to say here, a slice of 5D Minkowski space-time. So, the limiting perturbations of the induced metric on the brane $g_{\mu \nu}^{(4)}$, defined by
equation (\ref{induite}), 
should be solely supported by perturbations of the embedding functions $X^A(x^\mu)$, with $g^{(5)}_{AB}$ of equation (\ref{induite}) given by the background cosmological metric (\ref{backmet}). One may think that such a situation can only occur for specific cases (see e.g. \cite{NASH}). As a consistency check, one should be able to verify that one can indeed embed any scalar perturbation of the brane induced metric of the limiting form (that is to say obeying linearized 4D Einstein's equations) in a unperturbed 5D Minkowski bulk when the background brane Ricci scalar vanishes. 
Let us now show that this is the case.

We start with a brane defined as a hypersurface of coordinates $X^A(x^\mu)$ in the space-time parametrized by the metric (\ref{backmet}). 
The background embedding function $X^A_{(B)}(x^\mu)$ are given by 
\beq
X^A_{(B)}(x^\mu) = \delta^A_\mu x^\mu,
\eeq
so that one can write at linear order the $X^A(x^\mu)$ as 
\beq  
X^0& \equiv & t + \delta t, \\
X^i &\equiv& x^i + \delta^{ik} \partial_k \delta x ,\\
X^5 &\equiv& \delta y,
\eeq
where we have considered only embedding functions generating scalar perturbations on the brane world-volume. Defining then the linearization of the brane induced metric as 
\beq \label{linepertind}
g^{(4)}_{\mu \nu} = \left(\begin{array}{cc}
-(1+2A^{(4)}) & a_\bb^2 \partial_i B^{(4)} \\
a_\bb^2\partial_i B^{(4)} & a_\bb^2\left[(1+2 {\cal R}^{(4)})\delta_{ij}+ 2 \partial^2_{ij} E^{(4)}\right]
\end{array}\right),
\eeq
equation (\ref{induite}) leads to the following expressions for $A^{(4)}$, $B^{(4)}$, $R^{(4)}$ and $E^{(4)}$ as functions of $\delta t$, $\delta x$ and $\delta y$
\beq \nonumber
A^{(4)} &=& \dot{\delta t} + \frac{\dot{n}}{n} \delta t +
\frac{n^\prime}{n} \delta y, \\ \nonumber
{\cal R}^{(4)}&= &   \frac{\dot{a}}{a} \delta t +
\frac{a^\prime}{a} \delta y, \\ \nonumber
B^{(4)}&=&  -\frac{n^2}{a^2} \delta t + \dot{\delta x},\\ \nonumber
E^{(4)}&=&   \delta x,\\ \nonumber
\eeq
where the functions $a$ and $n$ appearing in those expressions are given by those of equations (\ref{a}-\ref{n}) evaluated in $y=0$.   If one then insists that the induced metric has the form (\ref{4Dlinepert}), one sees that one should choose $\delta x$ and $\delta t$ to vanish,   and $\delta y$ to obey the two equations 
\beq
\Phi &=& 
\frac{n^\prime}{n} \delta y, \label{dY1}
\\
\Psi &=& -\frac{a^\prime}{a} \delta y. \label{dY2}
\eeq
We wish in addition to impose $\Phi$ equal to $\Psi$, but $\Phi$ otherwise arbitrary. We see from the system (\ref{dY1}-\ref{dY2}) that this is possible only if the coefficient multiplying $\delta y$ in 
\beq
\Phi-\Psi = \left( \frac{n^\prime}{n} + \frac{a^\prime}{a}\right) \delta y,
\eeq
vanishes identically\footnote{Note further that whenever $H$ is non zero the same is true for $a^\prime/a$, so that a non vanishing $\delta y$ can indeed generate a non vanishing $\Psi$ perturbation, in contrast to the case of 
 a Minkowski brane.}. That is to say if   
\beq
R^\Bak_\bb/6 = \dot{H}+ 2 H^2 = 0, 
\eeq
where, to obtain this expression, we have used equations (\ref{a}) and (\ref{n}). Thus we conclude that arbitrary scalar perturbations of the brane can be supported only by fluctuations of the brane position if and only if the background Ricci scalar vanishes. This is then fully consistent with the calculations done in the first part of this subsection.

\subsection{Initial conditions and the limiting theory}
In the discussion carried in the previous subsections \ref{vDVZ2} and \ref{vDVZ4}, we considered sets of initial conditions on ${\cal C}_{(I)}$ chosen in a subset of all possible initial conditions, namely sets of the kind (\ref{SETIN}). Those initial conditions were chosen in particular so that they
 lead to a well behaved master variable $\Omega_{\cal S}^{\Upsilon \rightarrow 0}$ on the brane.   One could however worry that
 choosing a less restricted class of initial conditions could result in a limiting theory different from 4D linearized Einstein gravity. We would like to argue here that this is not the case unless the limiting $\Omega$ (or some of its derivatives) is diverging, leading to a subsequent divergence of bulk metric pertubations. Indeed, one can show that $\Omega$ fulfills the following equalities on the brane \cite{Deffayet:2002fn}
\beq
\Delta^2 \Omega_\bb  &=& -6 a^5 \left\{3 H \left( \dot{\Psi} + H \Phi \right) - a^{-2}\Delta \Psi + \eta \Upsilon^{-1}\kqd \delta \rho\right\}_\bb \label{Omat},\\
\Delta \Omega^\prime_\bb &=& -a^3 H^{-1}\left\{ a^{-3}H^2 \Delta \Omega -
2 \eta \Upsilon^{-1} \kqd \left(\delta \rho - 3 H \delta q - \Delta \delta \pi
\right)
 \right\}_\bb \label{Opmat},
\eeq
where we have defined $\delta \rho$, $\delta q$, $\delta \pi$ and (for future use) $\delta P$ as
 \beq \label{Er}
\delta \rho &=& \delta \rho_\MM + \frac{1}{\kqd}\left\{ 6 \frac{\dot{a}}{a}\left(\dot{ \Psi} + \frac{\dot{a}}{a} \Phi \right) - 2\frac{\Delta}{a^2} \Psi \right\}_\bb, \\ \label{Eq}
\delta q &=& \delta q_\MM + \frac{2}{\kqd} \left\{ \frac{\dot{a}}{a} \Phi + \dot{\Psi} \right\}_\bb, \\ \label{Epi}
\delta \pi &=& \delta \pi_\MM + \left\{ \frac{\Phi - \Psi} {\kqd a^2} \right\}_\bb,\\
\label{Ep}
\delta P &=& \delta P_\MM - \frac{1}{\kqd} \left\{  \left( 4 \frac{\ddot{a}}{a} +2 \frac{\dot{a}^2}{a^2} \right) \Phi + 2 \frac{\dot{a}}{a} \dot{\Phi} + 2 \ddot{\Psi} + 6 \frac{\dot{a}}{a} \dot{\Psi} + \frac{2 \Delta }{3 a^2} (\Phi - \Psi) \right\}_\bb.
\eeq
On the other hand, $\Omega$ is expressed in terms of the gauge invariant scalar perturbations of the 5D metric, $\tilde{A}$, $\tilde{A}_y$, $\tilde{A}_{yy}$ 
(defined in equations (\ref{AGIV}-\ref{AYYGIV})) 
as \cite{Deffayet:2002fn} 
\beq 
\frac{2 \da \paBDL}{3 a^3} \Delta^2 \Omega &=&  \dot{a}a^\prime \left( 4 \Delta(\tilde{A} + \tilde{A}_{yy} ) + 6 \dot{a}^2_\bb ( \tilde{A} + 2 \tilde{A}_{yy} ) - 3 \dot{a}_\bb a^\prime \tilde{A}_y\right) \nonumber \\ &&
+ 6 a \dot{a}^2_\bb  \left(2 \ddot{a}_\bb \tilde{A}_y + \dot{a} \tilde{A}^\prime + 2 a^\prime \tilde{A}^\cdot \right) \nonumber \\
&& + 3 a \dot{a}_\bb a^\prime \left(2\ddot{a}_\bb(\tilde{A}-\tilde{A}_{yy}) + 2\dot{a}  \tilde{A}_y^\prime + a^\prime  \tilde{A}_y^\cdot \right),
\label{OmeGIV}
\eeq
which is holding everywhere in the bulk (and thus also on the brane). 
Notice then that the right hand side of 
equations (\ref{Er}-\ref{Ep}) are the linearized 4D Einstein equations,
 so that the non vanishing of $\delta \rho$, $\delta q$, $\delta \pi$ and $\delta P$ measures the failure of the perturbation theory to match with usual 4D linear cosmological perturbations theory. Let us now consider, e.g. equation (\ref{Omat}) in the limit $\Upsilon \rightarrow 0$. With a choice of initial conditions (\ref{SETIN}) and a background with a non vanishing scalar curvature, we saw previously that $\delta \rho$ vanishes as $\Upsilon$, when we let $\Upsilon$ go to zero. It is then easy to see from equation (\ref{Omat}) that the only possibility for $\delta \rho$ to have a non vanishing limit as $\Upsilon$ goes to zero, is if $\Delta^2 \Omega_\bb$ diverges simultaneously\footnote{We remind that in this limit, we always ask the 4D ``observables'' on the brane, such as $\Phi$ or $\dot{\Phi}$ to be bounded.}. Now if 
$\Delta^2 \Omega$ diverges on the brane, equation (\ref{OmeGIV}) shows that some of the gauge invariant variables (or some of their first derivatives) have to diverge as well, leading to a breakdown of the linearized perturbation theory. 
A similar reasoning can be made starting from equation (\ref{Opmat}), or in the case of a background with vanishing Ricci scalar. 
\label{vDVZ5}

\section{Breakdown of the linearized theory}
So far we have only discussed the linearized DGP model.
Namely, we have rephrased the question of the presence or absence of the vDVZ discontinuity over cosmological background into that of knowing what is the limiting theory, as $r_cH \rightarrow \infty$ of linear DGP cosmological perturbations theory, when such a limit exists. This is a question one can ask whatever the validity range of the linear theory is. We would like now to question this validity range. As we will see, this can shed also an interesting light on the vDVZ discontinuity, and in particular on the difference between the cases with vanishing and non-vanishing background Ricci scalar. 
For reasons that will become clear in the following, and motivated by a similar discussion held for the metric around spherically symmetric bodies \cite{Tanaka:2003zb,Nicolis:2004qq}, we will concentrate on the brane extrinsic curvature $K_{\mu \nu}^{(b)}$. The latter is expressed in terms of the brane effective energy momentum tensor $S_{\mu \nu}$ by the 
Israel's junction conditions \cite{Israel}, derived from equation 
(\ref{einstein}) 
\beq \label{IsraelTER}
K_{\mu \nu}^\bb = -\kcd \left(\frac{1}{2}  S_{\mu \nu} - \frac{1}{6} S g^{(4)}_{\mu \nu} \right),
\eeq
where the brane effective energy-momentum tensor is given by 
\beq \label{SDEF}
S_{\mu \nu} = T_{\mu \nu}^\MM - \frac{1}{\kqd} G^{(4)}_{\mu \nu}.
\eeq
Note that this tensor vanishes identically if the 4D Einstein equations are obeyed on the brane. Inserting this equation in the trace of equation (\ref{IsraelTER}) gives 
\beq \label{premiere}
K^\bb = \frac{\kcd}{6}\left(T_\MM + \frac{R_\bb}{\kqd} \right),	
\eeq
where $K \equiv K_{\mu \nu} g^{\mu \nu}_{(4)}$, $T_\MM \equiv T^\MM_{\mu \nu} g^{\mu \nu}_{(4)}$, and $R_\bb \equiv - G_{\mu \nu}^{(4)} g^{\mu \nu}_{(4)}$ is the Ricci scalar of the induced metric. Moreover, a Gauss decomposition of the 5D Einstein tensor leads to the equation (see e.g. \cite{BDL}) 
\beq \label{seconde}
R^\bb = K^{\bb}K_\bb - K^\bb_{\mu \nu} K_\bb^{\mu \nu}.
\eeq
Extracting the expression of $R_\bb$ from equation (\ref{premiere}) and inserting it into equation (\ref{seconde}), one finds the equation 
\beq \label{troisieme}
\frac{3}{r_c} K^\bb - K^{\bb}K_\bb + K^\bb_{\mu \nu} K_\bb^{\mu \nu} = \kqd T_\MM,
\eeq
which relates the brane extrinsic curvature to the brane matter sources.
Note that this last equation is valid in full generality.
Let us now consider truncating the theory to linearized perturbations. In this case one can decompose the extrinsic curvature tensor as 
\beq
K_{\mu \nu} = K_{\mu \nu}^\Back + K_{\mu \nu}^\Lin,
\eeq
where $K_{\mu \nu}^\Back$ is the background value of $K_{\mu \nu}$, and 
$K_{\mu \nu}^\Lin$, the piece of the extrinsic curvature linear in the metric perturbations. Linearizing equation (\ref{troisieme}) over the background metric, one obtains  
\beq \label{quatrieme}
\frac{3}{r_c} K^\bb_\Lin + \tilde{K}_\Lin = \kqd T_\MM^\Lin, 
\eeq
where $\tilde{K}_\Lin$ is defined by
\beq
\tilde{K}_\Lin =- 2 K^{\bb}_\Lin K^\bb_\Back + K^{\bb \Back}_{\mu \nu} K_{\bb \Lin}^{\mu \nu} + K^{\bb \Lin}_{\mu \nu} K_{\bb \Back}^{\mu \nu},
\label{cinquieme}
\eeq
and $K_\Lin$ is linear into the metric perturbations. 
Notice that if one goes beyond the linearized theory, the left hand side of equation (\ref{troisieme}) contains necessarily the term 
\beq \label {NONLIN}
K^{\bb \Lin}_{\mu \nu} K_{\bb \Lin}^{\mu \nu} - K^{\bb}_\Lin K^\bb_\Lin.
\eeq
Let us first discuss the case of a Minkowski background. In this case $K_{\mu \nu}^{\Back}$ is vanishing so that one expects the linear perturbation theory to break down at least when the terms (\ref{NONLIN}), quadratic in the linearized extrinsic curvature, become of the same order as the first term on 
the left hand side of equation (\ref{quatrieme}). 
 This is what is found in references \cite{Tanaka:2003zb,Nicolis:2004qq} for the case of a non relativistic point like source. The breaking down of the 
linear perturbation theory can thus be estimated to happen when 
\beq
K^{\bb}_\Lin K^\bb_\Lin \sim K^\bb_\Lin / r_c, 
\eeq
 or 
\beq \label{CONDLINE}
K^{\bb}_\Lin r_c \sim 1.
\eeq
In the case of a non relativistic point like source, we note that equations (\ref{IsraelTER}) and (\ref{SDEF})
show that $K^\Lin_{\mu \nu}$ is of the order\footnote{One could have thought that $S_{\mu \nu}$ is vanishing as $r_c^{-1}$, when one lets $r_c$ go to infinity. This is however not so. Indeed, because there is a vDVZ discontinuity on a Minkowski background, the 4D Einstein equations are not fulfilled in the limit of large $r_c$, and there remains a non zero contribution in $S_{\mu \nu}$.} of $r_c  G^{(4) \Lin}_{\mu \nu}$. We should then 
have for dimensional reasons 
$K^\Lin_\bb \sim r_c r_S r^{-3}$, where $r_S$ is the (usual) Schwarzschild radius of the source, and $r$ is the distance to the source.  Thus, from equation (\ref{CONDLINE}) one concludes that the linearized theory is no longer valid 
for distances to the source smaller than $r_v$,
$r_v$ being given by equation (\ref{strongcoupl}).
This is what is found in ref. \cite{Deffayet:2002uk,Lue:2001gc,Gruzinov:2001hp,Porrati:2002cp,Tanaka:2003zb,Nicolis:2004qq} in agreement with the similar behavior found in the Pauli-Fierz theory \cite{Arkady}. Notice that $r_v$ goes to infinity as $r_c$ goes to infinity. This indicates in particular that, for a fixed source, and at a given distance to this source, the linear perturbation theory is never a good approximation as $r_c$ tends to infinity. 
\label{VDVZFRWPAR28}

Let us now discuss the case of a cosmological background, and concentrate on the kind of non linearities (\ref{NONLIN}) that are responsible for the appearance of the ``Vainshtein'' radius $r_v$ as recalled above. In this case, $K^\Back_{\mu \nu}$ is no longer vanishing. Rather, from the identity 
\beq
K^\bb_{\mu \nu} = \left\{\frac{1}{2} \partial_y g_{\mu \nu}^{(4)}\right\}_{y=0},
\eeq
valid in a Gaussian Normal gauge, where the 5D metric takes the form (\ref{GN}) and the brane sits in $y=0$, as well as from equations (\ref{a}-\ref{n}), one finds the non vanishing components of $K_{\mu \nu}^\Back$ to be given by
\beq
K^{\bb \Back}_{00} &=& \eta \frac{\dot{H}+H^2}{H}, \\
K^{\bb \Back}_{ij}  &=& - \eta H a^2_\bb \delta_{ij},
\eeq
so that $K^\Back_\bb$ is of the order $H$.
This means in particular that $\tilde{K}_\Lin$ is likely to dominate over the first term in the left hand side of equation (\ref{quatrieme}), when one considers the limit $\Upsilon \rightarrow 0$.  
This can indeed be checked as follows. We want first to estimate the components of $K_{\mu \nu}^\Lin$ in term of $\Omega$ and its derivatives, as well as their dependence in $r_c$. To do so, we work the GN gauge (\ref{GN}), where the 5D perturbed metric (\ref{linepertzero})
is given by 
\beq \label{linepert}
g_{AB} = \left(\begin{array}{ccc}
-n^2(1+2A) & a^2 \partial_i B & 0 \\
a^2\partial_i B & a^2\left[(1+2 {\cal R})\delta_{ij}+ 2 \partial^2_{ij} E\right]&0\\
0 & 0  &1
\end{array}\right),
\eeq
and we have kept only scalar metric perturbations, $A, B, {\cal R},$ and $E$\footnote{So that $A, B, {\cal R},$ and $E$ are the values of  $\bar{A}, \bar{B}, \bar{\cal {R}},$ and $\bar{E}$ in the gauge considered}.
The following gauge conditions can be further imposed without loss of generality
\beq 
E_\bb &=& 0, \\
B_\bb &=& 0,  
\eeq
and the induced metric on the brane is given by the line element (\ref{4Dlinepert}) with $\Phi$ and $\Psi$ given by 
\beq
A_\bb&=& \Phi, \\ 
{\cal R}_\bb&=&- \Psi. \\
\eeq
Israel's junction conditions (\ref{IsraelTER}) translates into the following set of equalities \cite{Deffayet:2002fn}
\beq 
{A}^\prime_\bb &=& \frac{\kcd}{6}\left(3 \delta P + 2 \delta \rho\right) \label{DPdef}, \label{Ap} \\
{{\cal R}}^\prime_\bb & =& \frac{1}{6} \kcd \left( \Delta \delta \pi - \delta \rho \right), \label{Rpdef} \\
{B}^\prime_\bb &=& \kcd \frac{n_\bb^2}{a_\bb^2} \delta q \label{Dddef}, \\
{E}^\prime_\bb &=& -\frac{1}{2} \kcd \delta \pi. \label{Epri}
\eeq
Using those formulae, as well as equations (\ref{PHI1}-\ref{DP1}), it is easy to extract the $r_c$ dependence of the linearized extrinsic curvature $K^\bb_\Lin$ (as well as for $K_{\mu \nu}^{\bb \Lin}$ and  $K^{\mu \nu}_{\bb \Lin}$). 
Let us first assume, as previously that the background Ricci scalar
does not vanish on some time interval $[t_0,t_{max}]$. One finds then 
 that, at the leading order as $\Upsilon$ goes to zero,  $K^\bb_\Lin$
is given by a sum of terms of the form 
\beq \label{ExpressKL}
 \alpha_{(p,q,r)}  \partial^q_y \partial_t^r \Delta^p \Omega, 
\eeq
where $p \in \{0,1,2\}$, $q \in \{0,1\}$, and $r \in \{0,1,2,3,4\}$, and 
$\alpha_{(p,q,r)}$ are time dependent coefficients depending only on the background metric through the scale factor $a$ and its time derivatives. 
Notice in particular that the linearity of $K_{\mu \nu}$ in $r_c$ (through $\kcd$), implied by equation (\ref{IsraelTER}) (or equations (\ref{Ap}-\ref{Epri})), has disappeared.  This is because the right hand side of equation (\ref{SDEF}) is found also to depend linearly on $r_c^{-1}$ at leading order, when considering a cosmological background; or, to say it in an other way, as we saw in section \ref{vDVZ2}, because the right hand sides of equations (\ref{Er}-\ref{Epi}) vanish as 
$\Upsilon$, as we let $\Upsilon$ go to zero and keep $\Omega$ (and its derivatives) bounded. 
This contrasts with the case of a flat Minkowski background, as seen above.
Notice also that equations (\ref{DR1}-\ref{DP1}) lead to a form similar 
to that of $K^\bb_\Lin$ for $T_\MM^\Lin$. We know in addition that the limiting expression of $T_\MM^\Lin$, as $\Upsilon$ goes to zero, can be made non vanishing, this because it is obtained solving the 4D linearized Einstein equations (\ref{EQM6_4D}-\ref{EQM8_4D}). While in the same limit, we conclude from the form (\ref{ExpressKL}) that $K^\bb_\Lin/r_c$ is in fact vanishing. 
so that the dominant term in the left hand side of equation (\ref{quatrieme}) is $\tilde{K}_\Lin$ of order  $K^\bb_\Lin H$. 
The next step is then to ask whenever the nonlinear term (\ref{NONLIN}) become of the same order as the dominant linear ones. This happens now when 
\beq
K^\bb_\Lin H^{-1} \sim 1, 
\eeq
which does not contain $r_c$. This means that when $\Upsilon$  goes to zero, one should find a modified, $r_c$-independent, Vainshtein scale\footnote{I Thank Arthur Lue for discussing this point.}. This is indeed what has been found in references \cite{SOLAR,Lue:2004rj}. 
The same reasoning and conclusion would be true in the case of a background with a vanishing Ricci scalar on the time interval $[t_0,t_{max}]$. In this case, the preceding discussion is unchanged once one replaces $\Omega$ by $\tilde{\Omega}$.

Last, we would like to ask what is happening if one considers a background with a Ricci scalar that vanishes only on some set of zero measure. Consider, e.g., a background such that its Ricci scalar does not vanish on some interval $[t_0, t_{max}[$, but vanish at the end of it, in $t_{max}$. Then, fix some initial data
 in $t_0$ and in the bulk, following the procedure explained in section \ref{vDVZ2}, and consider the limit $H r_c \rightarrow \infty$. If one does so, one will find that the master variable $\Omega$, as well as its derivatives, stay perfectly well behaved in the bulk\footnote{disregarding some possible singularities arising due the hyperbolic nature of the problem.}. This is then also true for the bulk gauge invariant variables $\tilde{A}$, $\tilde{A}_y$, $\tilde{A}_{yy}$ and $\tilde{\cal R}$ (see equations (\ref{AGIV}-\ref{RGIV})), and so also for the bulk geometry. On the other hand, the 4D observable quantities on the brane, $\Phi, \Psi, \delta \rho_\MM, \delta q_\MM,$ and $\delta P_\MM$ have in general a bounded (and continuous) limit for $t < t_{max}$, but diverge in $t_{max}$. This is because of the divergence, underlined in section (\ref{vDVZ4}), of the coefficients ${\cal C}$, as $\Upsilon$ and $R^B_\bb$ go to zero.  
The same derivation that lead to (\ref{ExpressKL}) shows that $K^{\bb \Lin}_{\mu \nu}$ is also diverging in $t=t_{\max}$.
One thus concludes that the linearized theory is breaking down 
when one approaches this moment. A similar conclusion is reached if one considers the gauge invariant brane location $\tilde{\xi}$. As can be seen from equations (\ref{POSPHIPSI}) and (\ref{PHI1}-\ref{PSI1}), the limiting $\tilde{\xi}$ is finite for $t<t_{max}$, but
 diverges at $t_{max}$. 
Those divergences constitute somehow a reappearance of the vDVZ discontinuity
when one goes continuously from a background with non vanishing Ricci scalar to a background with a vanishing one. They are of course likely to be absent in the full non linear theory. 

\label{SECQUAT}

\section{Discussion and Conclusion}
In this paper, we discussed the theory of linear cosmological perturbations 
of the Dvali-Gabadadze-Porrati model. We distinguished the case of a brane background space-time with a non vanishing Ricci scalar from that of a brane background space-time with a vanishing Ricci scalar. In the first case, we showed that 
if one fixes at some initial time along the brane world-volume, the initial values of the gravitational potential $\Phi$ and its time derivative $\dot{\Phi}$, as can be done in 4D general relativity, and if one further insists that the initial data to be provided in the bulk have a well-behaved (continuous and bounded) limit when we let the radius $r_c$ of transition between the 4D and 5D regime go to infinity with respect to the Hubble radius of the background space-time, then the linearized theory on the brane does have a limit, and this limit is given by the theory of cosmological perturbations derived from standard 4D GR.
This holds for the simplest case of matter living on the brane, namely a perfect fluid with adiabatic perturbation, or a scalar field. This result can be rephrased in saying that the boundary condition on the brane, derived from the matter equation of state, and that is usually ``non local'' in time \cite{Kodama:2000fa}, degenerates, in the limit considered, into a standard Dirichlet boundary condition that is obtained solving the linearized 4D Einstein's equations. So, in fact, in this limit, the brane decouples from the bulk. It contrasts with the case of a Minkowski background, 
and can be interpreted as showing, in a ``Hamiltonian'' sense, that the vDVZ discontinuity does disappear on a FLRW background  in a similar way as it does on a maximally symmetric background with a non vanishing curvature.
What we mean here by that, is that an observer living on the brane, and considering only scalar perturbations, thus, say, fixing some initial data as he would do in standard 4D cosmology, will not be able to distinguish the time evolution of these data in the DGP model with large enough $r_c H$, from what he would have observed in standard cosmology with the same initial data.
The same result is found to hold in the case of a brane background space-time with a vanishing Ricci scalar, with however the important difference that the initial data in the bulk are to be such that their limiting values vanish. This contrast with the previous case where the only constraint those bulk initial data have to fulfill are those implied by the choice of initial data on the brane.
Thus, in the case of a brane with a vanishing Ricci scalar, the limiting perturbations are  solely supported by the brane motion in a bulk empty from metric perturbations.
Moreover, we have also argued that the vDVZ discontinuity somehow reappears on a cosmological background where the Ricci scalar does not vanish but on a measure zero subset. In this case, if one only considers the time evolution of perturbations given by the linearized theory and follows the procedure outlined above for fixing initial conditions, one finds divergences in observable quantities at the points where the Ricci scalar vanishes when one lets $r_c H$ go to infinity. These divergences result in a breaking down of the linearized theory, in a similar way to what is happening around a static source on a 
Minkowski background \cite{Deffayet:2002uk,Lue:2001gc,Gruzinov:2001hp,Porrati:2002cp,Tanaka:2003zb}, which is intimately related to the vDVZ discontinuity.  One might be concerned that our results seem to depend on the assumption the bulk initial data have a continuous and bounded limit. In fact, the only necessary condition is that those bulk initial data are such that the limiting value (as $r_c H$ goes to infinity) on the brane of the master variable and its derivatives (or of $\tilde{\Omega}$ in the case of a vanishing $R^{(B)}_{(b)}$) are bounded\footnote{or if they are not, that the limiting 4D observable expressions $\Phi^{\Upsilon \rightarrow 0}$, $\Psi^{\Upsilon \rightarrow 0}$, $\delta \rho^{\Upsilon \rightarrow 0}_\MM$, $\delta P^{\Upsilon \rightarrow 0}_\MM$ and $\delta q^{\Upsilon \rightarrow 0}_\MM$ are bounded.}, as seen in particular in section \ref{vDVZ5}.
When it is not the case, we argued that the linear perturbation theory is no longer valid.

\label{SECCINQ}

\section*{Acknowledgments}
We thank G.~Dvali, T.~Damour, G.~Esposito-Farese, G.~Gabadadze, K.~Koyama, A.~Lue, K.~Malik, J.~Mourad, M.~Porrati, R.~Scoccimarro, J.~Shata, T.~Tanaka,  and M.~Zaldarriaga for useful discussions, as well as D.~Langlois for the same and a careful reading of this manuscript.

\appendix

\section{Expression of the coefficients $\CC$}\label{expressC}

Here we give the expressions for the coefficients appearing in equations (\ref{PHI1}-\ref{DP1}). 
\beq
\CC^\Phi_{\Delta (0,0)} &=& 1+ \fU (\UU + 1) \nonumber, \\
\CC^\Phi_{ (1,0)} &=& 3 \{ 1+ \fU (\UU + 1)\}  \nonumber,\\
\CC^\Phi_{ (2,0)} &=& -3\fU (1+ \Upsilon) \nonumber, \\
\CC^\Phi_{ (0,1)} &=&  3 \eta  \{  \fU (\UU+2)( \UU+1) -\UU \}\nonumber ,\\
\CC^\Psi_{\Delta (0,0)}&=& 1 + \fU \nonumber, \\
\CC^\Psi_{ (1,0)}&=& 3(1+ \fU)  \nonumber,\\
\CC^\Psi_{ (2,0)}&=& - 3 \fU \nonumber,\\
\CC^\Psi_{ (0,1)}&=& 3 \eta \fU (2+\UU)\nonumber,\\
\CC^\rho_{\Delta (0,0)} &=& \frac{\fU^2}{H^3}  \left( 2 + \UU  \right) \left\{ 12 H^3 + 10 \dot{H} H + \ddot{H} + \UU  \left( 8 H^3  + 2 \dot{H} H \right) + H^3 \UU^2 \right\}\nonumber ,\\
\CC^\rho_{\Delta^2 (0,0)} &=& \frac{\fU}{3 H^2}  \left(  6H^2 + 2 \hd  +3 H^2 \UU \right)\nonumber,\\
\CC^\rho_{(1,0)} &=& - \frac{3 \fU^2}{H^6} \left( 2 + \UU \right) \left\{ 
6 H^4 \hd + 7 H^2 \hd^2 + \hd^3 - H^3 \hdd  \right.\nonumber\nonumber \\
&&+ \left. \UU \left( 2 H^6 + 7H^4 \hd + 2 H^2 \hd^2 \right) + \UU^2 \left( H^6 + H^4 \hd \right) \right\}\nonumber,\\
\CC^\rho_{(2,0)} &=& - \frac{\fU^2}{H^4} \left( 2 + \UU \right)  \left\{ 3\left(6 H^4 + 5 H^2 \hd - \hd^2 + H \hdd \right) + 9H^4 \UU \right\}\nonumber, \\
\CC^\rho_{\Delta (2,0)} &=& - \fU \left( 2 + \UU \right)\nonumber, \\
\CC^\rho_{(3,0)} &=& 3  \fU  \left( 2 + \UU \right)\nonumber, \\
 \CC^\rho_{(0,1)} &=& -\eta \frac{3 \fU^2}{H^4} \left( 2 + \UU \right) \left\{  
4 H^2 \hd + 6 \hd^2 - 2 H \hdd + \UU \left( 2 H^2 \hd + \hd^2 - H \hdd \right)\right\} \nonumber, \\ \nonumber
\CC^\rho_{\Delta (0,1)} &=&    \eta \fU \left( 2 + \UU \right)^2\nonumber,\\
\CC^\rho_{ (1,1)} &=& -3  \eta \fU \left( 2 + \UU \right)^2 \nonumber,\\
\CC^q_{\Delta(0,0)} &=& \frac{\fU^2}{H^4} \{24 H^4 + 20 H^2 \hd + 2 H \hdd + \UU (24 H^4 + 10 H^2 \hd - \hd^2 + H \hdd) + 6 \UU^2  H^4 \},\nonumber  \\  
\CC^q_{(1,0)} &=& -\frac{3 \fU^2}{H^6} (2+\UU)\{ \UU (2+ \UU) H^6 + (6 + 7 \UU + \UU^2 ) H^4 \hd + (7+2 \UU) H^2 \hd^2 + \hd^3 - H^3 \hdd\} \nonumber, \\
\CC^q_{\Delta(1,0)} &=& -\frac{\fU}{H^2} \{3H^2 (2+ \UU) +2 \hd\}  \nonumber, \\ 
\CC^q_{(2,0)} &=&  -\frac{3 \fU^2}{H^4} (2+\UU) \{ 3 (2+\UU) H^4 + 5 H^2 \hd - \hd^2 + H \hdd \} \nonumber, \\
\CC^q_{(3,0)} &=&  3 \fU (2+\UU)  \nonumber, \\
\CC^q_{(0,1)} &=&  -\eta \frac{3 \fU^2}{H^4} (2+\UU)\{ 2 (2+ \UU) H^2 \hd + (6 + \UU) \hd^2 - (2+ \UU) H \hdd \} \nonumber, \\
\CC^q_{(1,1)} &=& -3 \eta  \fU ( 2 + \UU)^2, \nonumber \\
\CC^P_{\Delta(0,0)} &=& \frac{\fU^3}{H^8}\left\{ (2+\UU)^3 (-3+2\UU)H^6 \hd + (2+\UU) \hd^4 + 2 (2+\UU)^2 H^5 \hdd \nonumber \right.\\  && \left.
+ (28+20 \UU + 3 \UU^2) H^3 \hd \hdd - (4+ \UU) H \hd^2 \hdd + (2 + \UU) H^4 ((-24+ 5 \UU^2) \hd^2 \nonumber \right.\\  && \left.
- (2+ \UU) \hddd) - H^2 ((46+13 \UU - 2 \UU^2) \hd^3 - 2 (2+\UU) \hdd^2 + (2+ \UU) \hd \hddd) \right\} \nonumber, \\
\CC^P_{(1,0)} &=& \frac{\fU^3}{H^{10}}\left\{ 3 (2 \UU(2+ \UU)^3 H^{10} + (2 + \UU)^2(18 +25 \UU + 5 \UU^2) H^8 \hd - 3(-6 +\UU^2) H^2 \hd^4 \nonumber \right.\\  && \left.-\UU \hd^5 + (2+\UU)^2 (4 + 5\UU + \UU^2) H^7 \hdd + (44+52 \UU + 21 \UU^2 + 3 \UU^3) H^5 \hd \hdd \nonumber \right.\\  && \left. + (6+ 10 \UU + 3 \UU^2) H^3 \hd^2 \hdd + (2+\UU) H \hd^3  \hdd - 
(2+ \UU) H^6 ((-56-56\UU - 10 \UU^2 + \UU^3) \hd^2 \nonumber \right.\\  && \left. + (2+ \UU) \hddd) + H^4((42+ 39 \UU + \UU^2- 3 \UU^3) \hd^3 + 2(2+ \UU) \hdd^2 - (2+ \UU) \hd \hddd))\right\}_\bb \nonumber ,\\
\CC^P_{\Delta (1,0)} &=& -\frac{2 \fU^2}{ H^4} 
( 2 + \UU )\left\{ 3 \left( 2 + \UU \right) H^4 + 5 H ^2 \hd - \hd^2 + H \hdd  \right\}_\bb \nonumber, \\
\CC^P_{(2,0)} &=& \frac{\fU^3}{ H^8} \left\{3((2+ \UU)^3(6+\UU) H^8 + 2 (2+ \UU)^2 (13+4 \UU) H^6 \hd + \UU \hd^4\nonumber \right.\\  && \left. - 3(8+6 \UU + \UU^2) H^3 \hd \hdd + (4+ \UU) H \hd^2 \hdd
+ (2+ \UU) H^4 ((42+19\UU+ \UU^2) \hd^2\nonumber \right.\\  && \left. + (2+\UU) \hddd) + H^2 ((54+ 30 \UU+ 4 \UU^2) \hd^3 - 2 (2+ \UU) \hdd^2 + (2+ \UU) \hd \hddd))\right\}_\bb ,
\nonumber \\
\CC^P_{\Delta (2,0)} &=& \frac{\fU}{H^2} \left\{ 3(2+ \UU) H^2 + 2 \hd \right\}_\bb \nonumber ,\\
\CC^P_{ (3,0)} &=&  \frac{3 \fU^2}{H^4} \left\{ (2+ \UU)^2 H^4 + 3(2+ \UU) H^2 \hd - 2 (3+ \UU) \hd^2 + 2 (2+ \UU) H \hdd \right\}_\bb \nonumber, \\
\CC^P_{ (4,0)} &=& -3 \fU (2+ \UU) \nonumber, \\
\CC^P_{(0,1)} &=&  \eta \frac{3\fU^3}{H^8} \left\{(2+ \UU)^3 (6+ \UU) H^6 \hd + (24+ 6 \UU - \UU^2) \hd^4 + (2+ \UU)^2(10+ 3 \UU) H^3 \hd \hdd\nonumber \right.\\  && \left. - (12+ 8 \UU + \UU^2) H \hd^2 \hdd - (2+ \UU)^2 H^4 ((-20 -2\UU + \UU^2) \hd^2 + (2+\UU) \hddd )\nonumber \right.\\  && \left. - (2+ \UU) H^2 ((-6+ 9 \UU + 4 \UU^2) \hd^3 - 2(2+ \UU) \hdd^2 + (2+ \UU) \hd \hddd) \right\}_\bb\nonumber ,\\
\CC^P_{(1,1)} &=& \eta \frac{6 \fU^2}{H^4} (2 + \UU) \left\{(2+ \UU)^2 H^4 + 3(2+\UU) H^2 \hd + (6+\UU) \hd^2 - (2+ \UU) H \hdd \right\} \nonumber, \\
\CC^P_{(2,1)} &=&  3 \eta  \fU (2+ \UU)^2, \nonumber 
\eeq
where $\hddd$ denotes the third derivative of $H$ with respect to the brane cosmic time $t$, and we have defined $\fU$ by 
\beq
\fU &=& \left(\Upsilon + \frac{R^\Bak_\bb}{6H^2} \right)^{-1}, \nonumber
\eeq
${R^\Bak_\bb} =  6 \left(\dot{H}+2 H^2 \right)$ being the background scalar curvature of the induced metric on the brane.


\begin{thebibliography}{99} 

\bibitem{Arkani-Hamed:1998rs}
N.~Arkani-Hamed, S.~Dimopoulos and G.~R.~Dvali,
Phys.\ Lett.\ B {\bf 429} (1998) 263
[arXiv:hep-ph/9803315].
I.~Antoniadis, N.~Arkani-Hamed, S.~Dimopoulos and G.~R.~Dvali,
Phys.\ Lett.\ B {\bf 436} (1998) 257
[arXiv:hep-ph/9804398].
N.~Arkani-Hamed, S.~Dimopoulos and G.~R.~Dvali,
Phys.\ Rev.\ D {\bf 59} (1999) 086004
[arXiv:hep-ph/9807344].

\bibitem{Randall:1999vf}
L.~Randall and R.~Sundrum,
Phys.\ Rev.\ Lett.\  {\bf 83} (1999) 4690
[arXiv:hep-th/9906064].

\bibitem{GRS}
C.~Charmousis, R.~Gregory and V.~A.~Rubakov,
Phys.\ Rev.\ D {\bf 62} (2000) 067505
[arXiv:hep-th/9912160].
R.~Gregory, V.~A.~Rubakov and S.~M.~Sibiryakov,
Phys.\ Rev.\ Lett.\  {\bf 84} (2000) 5928
[arXiv:hep-th/0002072].


\bibitem{Pilo:2000et}
L.~Pilo, R.~Rattazzi and A.~Zaffaroni,
JHEP {\bf 0007} (2000) 056
[arXiv:hep-th/0004028].

\bibitem{Rubakov:2001kp}
V.~A.~Rubakov,
Phys.\ Usp.\  {\bf 44} (2001) 871
[Usp.\ Fiz.\ Nauk {\bf 171} (2001) 913]
[arXiv:hep-ph/0104152].

\bibitem{DGP} G.~Dvali, G.~Gabadadze and M.~Porrati,
Phys.\ Lett.\  {\bf B485} (2000) 208
[arXiv:hep-th/0005016].

\bibitem{Deffayet:2001uy}
C.~Deffayet,
Phys.\ Lett.\ B {\bf 502}, 199 (2001)
[hep-th/0010186].

\bibitem{Fifth}
C.~Deffayet, G.~R.~Dvali and G.~Gabadadze,
Phys.\ Rev.\ D {\bf 65} (2002) 044023
[arXiv:astro-ph/0105068].


\bibitem{Deffayet:2002sp}
C.~Deffayet, S.~J.~Landau, J.~Raux, M.~Zaldarriaga and P.~Astier,
Phys.\ Rev.\ D {\bf 66}, 024019 (2002)
[arXiv:astro-ph/0201164].

\bibitem{SOLAR}
A.~Lue and G.~Starkman,
Phys.\ Rev.\ D {\bf 67}, 064002 (2003)
[arXiv:astro-ph/0212083].

\bibitem{SOLARBIS}
G.~Dvali, A.~Gruzinov and M.~Zaldarriaga,
Phys.\ Rev.\ D {\bf 68} (2003) 024012
[arXiv:hep-ph/0212069].

\bibitem{Veltman} H. van Dam and M. Veltman, Nucl. Phys. 
{\bf B22}, 397 (1970)~.
V.~I.~Zakharov, JETP Lett. {\bf 12}, 312 (1970)~.
 Y.~Iwasaki,
Phys.\ Rev.\ D {\bf 2} (1970) 2255.

\bibitem{Pauli}   M. Fierz, Helv. Phys. Acta 12 (1939) 3;
\\M. Fierz, W. Pauli, Proc. Roy. Soc. 173 (1939) 211.

\bibitem{Porrati:2000cp}
M.~Porrati,
Phys.\ Lett.\ B {\bf 498}, 92 (2001)
[arXiv:hep-th/0011152].


\bibitem{Higushi} A.~Higuchi,
Nucl.\ Phys.\ B {\bf 282}, 397 (1987); 
Nucl.\ Phys.\ B {\bf 325}, 745 (1989).

\bibitem{Deffayet:2002fn}
C.~Deffayet,
Phys.\ Rev.\ D {\bf 66} (2002) 103504
[arXiv:hep-th/0205084].


\bibitem{DEF}
C.~Deffayet,
arXiv:hep-th/0409302, to appear in Phys.~Rev.~D.


\bibitem{LICHNE}
A.~Lichnerowicz, ``Propagateurs et commutateurs en relativit\'e g\'en\'erale'', IHES, Publications Math\'ematiques, 10 (1961).



\bibitem{Kogan1} 
I.~I.~Kogan, S.~Mouslopoulos and A.~Papazoglou,
Phys.\ Lett.\ B {\bf 503}, 173 (2001)
[hep-th/0011138].

\bibitem{Deffayet:2002uk}
C.~Deffayet, G.~R.~Dvali, G.~Gabadadze and A.~I.~Vainshtein,
Phys.\ Rev.\ D {\bf 65} (2002) 044026
[arXiv:hep-th/0106001].

\bibitem{Arkani-Hamed:2002sp}
N.~Arkani-Hamed, H.~Georgi and M.~D.~Schwartz,
Annals Phys.\  {\bf 305}, 96 (2003)
[arXiv:hep-th/0210184].

\bibitem{Arkady} A.~I.~Vainshtein, Phys. Lett. 
{\bf 39B}, 393 (1972)~. 

\bibitem{BD}
D.~G.~Boulware and S.~Deser,
Phys.\ Rev.\ D {\bf 6}, 3368 (1972).

\bibitem{Isham:gm}
C.~J.~Isham, A.~Salam and J.~Strathdee,
Phys.\ Rev.\ D {\bf 3} (1971) 867.

\bibitem{Damour:2002ws}
T.~Damour and I.~I.~Kogan,
Phys.\ Rev.\ D {\bf 66} (2002) 104024
[arXiv:hep-th/0206042].

\bibitem{Duff:ea}
M.~J.~Duff, C.~N.~Pope and K.~S.~Stelle,
Phys.\ Lett.\ B {\bf 223} (1989) 386.


\bibitem{Porrati:2002cp}
M.~Porrati,
Phys.\ Lett.\ B {\bf 534}, 209 (2002)
[arXiv:hep-th/0203014].

\bibitem{Damour:2002gp}
T.~Damour, I.~I.~Kogan and A.~Papazoglou,
Phys.\ Rev.\ D {\bf 67} (2003) 064009
[arXiv:hep-th/0212155].

\bibitem{Gibbons:1976ue}
G.~W.~Gibbons and S.~W.~Hawking,
Phys.\ Rev.\ D {\bf 15}, 2752 (1977).

\bibitem{Lue:2001gc}
A.~Lue,
Phys.\ Rev.\ D {\bf 66}, 043509 (2002)
[arXiv:hep-th/0111168].

\bibitem{Gruzinov:2001hp}
A.~Gruzinov,
arXiv:astro-ph/0112246.

\bibitem{Tanaka:2003zb}
T.~Tanaka,
Phys.\ Rev.\ D {\bf 69} (2004) 024001
[arXiv:gr-qc/0305031].

\bibitem{Gabadadze:2004iy}
G.~Gabadadze and A.~Iglesias,
arXiv:hep-th/0407049.

\bibitem{DEBATE}
M.~A.~Luty, M.~Porrati and R.~Rattazzi,
JHEP {\bf 0309} (2003) 029
[arXiv:hep-th/0303116].
V.~A.~Rubakov,
arXiv:hep-th/0303125.
G.~Dvali,
arXiv:hep-th/0402130.
G.~Gabadadze,
arXiv:hep-th/0403161.

\bibitem{Nicolis:2004qq}
A.~Nicolis and R.~Rattazzi,
arXiv:hep-th/0404159.

\bibitem{BDL}
P.~Binetruy, C.~Deffayet and D.~Langlois,
Nucl.\ Phys.\ B {\bf 565} (2000) 269
[arXiv:hep-th/9905012].
P.~Binetruy, C.~Deffayet, U.~Ellwanger and D.~Langlois,
Phys.\ Lett.\ B {\bf 477} (2000) 285
[arXiv:hep-th/9910219].

\bibitem{Deruelle:2000ge}
N.~Deruelle and T.~Dolezel,
Phys.\ Rev.\ D {\bf 62} (2000) 103502
[arXiv:gr-qc/0004021].

\bibitem{Mukhanov:1992me}
J.~M.~Bardeen,
Phys.\ Rev.\ D {\bf 22} (1980) 1882.
V.~F.~Mukhanov, H.~A.~Feldman and R.~H.~Brandenberger,
Phys.\ Rept.\  {\bf 215} (1992) 203.
H.~Kodama and M.~Sasaki,
Prog.\ Theor.\ Phys.\ Suppl.\  {\bf 78} (1984) 1.


\bibitem{Mukohyama:2000ui}
S.~Mukohyama,
Phys.\ Rev.\ D {\bf 62}, 084015 (2000)
[arXiv:hep-th/0004067].


\bibitem{Bridgman:2001mc}
H.~A.~Bridgman, K.~A.~Malik and D.~Wands,
Phys.\ Rev.\ D {\bf 65} (2002) 043502
[arXiv:astro-ph/0107245].

\bibitem{Israel}
W.~Israel,
Nuovo Cim.\ B {\bf 44S10} (1966) 1
[Erratum-ibid.\ B {\bf 48} (1967\ NUCIA,B44,1.1966) 463].
G.~Darmois, M\'emorial des sciences math\'ematiques XXV (1927). 
K.~Lanczos, Ann. Phys. {\bf 74} (1924) 518. 
N.~Sen, Ann. Phys. {\bf 73} (1924) 365.

\bibitem{Giannakis:2002jg}
I.~Giannakis and H.~c.~Ren,
Phys.\ Lett.\ B {\bf 528} (2002) 133
[arXiv:hep-th/0111127].
R.~Dick,
Class.\ Quant.\ Grav.\  {\bf 18} (2001) R1
[arXiv:hep-th/0105320].

\bibitem{Kodama:2000fa}
H.~Kodama, A.~Ishibashi and O.~Seto,
Phys.\ Rev.\ D {\bf 62}, 064022 (2000)
[arXiv:hep-th/0004160].



\bibitem{courant}
R.~Courant, D.~Hilbert, ``Methods of Mathematical Physics'', Vol II, Interscience Publishers, 1962. 


\bibitem{VanNieuwenhuizen:1973qf}
P.~Van Nieuwenhuizen,
Phys.\ Rev.\ D {\bf 7} (1973) 2300.

\bibitem{NASH}
J.~Nash,
Annals of Mathematics, Vol~63 No.1 (1956) 20.


\bibitem{Lue:2004rj}
A.~Lue, R.~Scoccimarro and G.~D.~Starkman,
Phys.\ Rev.\ D {\bf 69} (2004) 124015
[arXiv:astro-ph/0401515].

\end{thebibliography}
\end{document}